\documentclass[a4paper,10pt]{article}
\usepackage{amsmath, amssymb, graphics,graphicx}
\usepackage{amscd}
\usepackage{textcomp}
\begin{document}
\title{Quantile Mechanics II:\\Changes of Variables in Monte Carlo methods\\and \\GPU-Optimized Normal Quantiles}
\author{William T. Shaw\thanks{Corresponding author: Departments of Mathematics and Computer Science, 
University College London, Gower Street, London WC1E 6B, England; E-mail: w.shaw@ucl.ac.uk}, Thomas Luu\thanks{Department of Mathematics, University College London; E-mail: t.luu@ucl.ac.uk} and Nick Brickman\thanks{Taylor Brickman Ltd; E-mail: nick@taylorbrickman.co.uk}}

\maketitle
\begin{abstract}
With financial modelling requiring a better understanding of model risk, it is helpful to be able to vary assumptions about underlying probability distributions in an efficient manner, preferably without the noise induced by resampling distributions managed by Monte Carlo methods. This article presents differential equations and solution methods for the functions of the form $Q(x) = F^{-1}(G(x))$, where $F$ and $G$ are cumulative distribution functions. Such functions allow the direct recycling of Monte Carlo samples from one distribution into samples from another. The method may be developed analytically for certain special cases, and illuminate the idea that it is a more precise form of the traditional Cornish-Fisher expansion.  In this manner the model risk of distributional risk may be assessed free of the Monte Carlo noise associated with resampling. The method may also be regarded as providing both analytical and numerical bases for doing more precise Cornish-Fisher transformations. Examples are given of equations for converting normal samples to Student t, and converting exponential to hyperbolic, variance gamma and normal. In the case of the normal distribution, the change of variables employed allows the sampling to take place to good accuracy based on a single rational approximation over a very wide range of the sample space. The avoidance of any branching statement is of use in optimal GPU computations as it avoids the effect of {\it warp divergence}, and we give examples of branch-free normal quantiles that offer performance improvements in a GPU environment, while retaining the best precision characteristics of well-known methods. We also offer models based on a low-probability of warp divergence. Comparisons of new and old forms are made on the  Nvidia Quadro 4000, GTX 285 and 480, and Tesla C2050 GPUs. We argue that in single-precision mode, the change-of-variables approach offers performance competitive with the fastest existing scheme while substantially improving precision, and that in double-precision mode, this approach offers the most GPU-optimal Gaussian quantile yet, and without compromise on precision for Monte Carlo applications, working twice as fast as the CUDA 4 library function with increased precision.
\end{abstract}
\bigskip
\noindent

\pagestyle{myheadings}
\thispagestyle{plain}
\markboth{W.T. Shaw}{W.T. Shaw, T. Luu \& N. Brickman:  Monte Carlo Changes of Variable}
\newpage

\section{Introduction}
The construction of Monte Carlo samples from a distribution is facilitated if one has a knowledge of the quantile function $w(u)$ of a distribution. If $F(x)$ is the cumulative distribution function, then the quantile $w(u)$ is the solution of the equation
\begin{equation}
F(w(u))=u\ . 
\end{equation}
A knowledge of the function $w(u)$ makes Monte Carlo simulation straightforward: given a random sample $U$ from the uniform distribution, a sample from the target distribution characterized by a density and distribution $f(x), F(x)$, is 
\begin{equation}
X = w(U)\ .
\end{equation}
While it is commonplace to use the uniform distribution on the unit interval as the base distribution for sampling, there is in fact no need to do so\footnote{This rather clear observation was first made to me by Peter Jaeckel}. In his critique of copula theory \cite{mikosch}, T. Mikosch stated {\it There is no particular mathematical or practical reason for transforming the marginals to the uniform distribution on (0; 1)} and proceeded to consider exponential and normal coordinates. On the other hand, samples from the uniform distribution are readily generated, and in the computer science domain it remains a topic of continuing study to develop better uniform random number generators. 

For example, a great deal of intellectual effort has been expended on highly efficient sampling from the normal and other well-known distributions. Given such samples, can we leverage the work done to create samples from other distributions in an efficient manner? This article will address this question in the affirmative. In principle the answer is trivial: given a sample $Z$ from a distribution with CDF $G(x)$, we first work out $G(X)$ which is uniform. Then we can apply the quantile function $F^{-1}(x)$ associated with a target distribution with CDF $F$ and form  $F^{-1}(G(x))$ as a sample from that target distribution. In general $F$, $G$ and their inverses can be rather awkward special functions (see e.g. \cite{shawjcf06}) , so a direct route to the object $Q(x) = F^{-1}(G(x))$ would be helpful. 

There are at least two ways of developing this idea. One route is to {\it postulate} interesting forms for the composite mapping. This has been explored by Shaw and Buckley \cite{imasb} based on Gilchrist's theory of quantile transformations \cite{warren}. In this way we can find skew and kurtotic variations of {\it any} base distribution, while avoiding, in a controlled manner, the introduction of ``negative density'' problems that arise in traditional Gram-Charlier methods. The second route is to try to simplify the mapping {\it given} a choice of $F$ and $G$. Such a route can be found by the method of differential equations for quantile functions developed by Steinbrecher and Shaw \cite{qmone}. In the next section we will give a brief review of that approach, and in Section 4 we shall use this insight to build a new representation of the Student t quantile in terms of a series expansion of a Gaussian. This neatly tidies up the well-known expansions already available for the Student distribution as the degrees of freedom becomes large, and we obtain a more compact power series in the normal random variable. 

In view of the importance of Gaussian sampling, we will also explore how to make normal random variables by the change of variables approach, and explore the use of the exponential and even Student distributions as intermediaries. In this simple cases we can write down the objects $F^{-1}(G(z))$ very explicitly and the issue becomes one of the efficient approximation and computation of such objects. We will show that this offers a useful performance benefit in a GPU environment, where branching algorithms may be subject to significant performance penalties. Our change-of-variables approach will allow costly branching to be avoided or reduced to a low probability and we will demonstrate the benefits in the CUDA environment for programming NVIDIA GPUs. This part of the paper of necessity involves ideas from computer science as much as applied mathematics, but there are lessons in approximation theory to be learnt from this analysis, which reveals the relative merits of using different types of error norm in the constructions in terms of their impact on the actual precision of the relevant function approximations. 

The plan of this work is as follows. In Section 2 we review the differential approach to quantile functions and introduce a new differential equation that characterizes the mapping from samples of one probability distribution to another. We give specific examples for recycling samples of exponential and Gaussian random variables. In Section 3 we focus on Gaussian backgrounds and review the transformation from Gaussian random variables to those from  Student t distribution. We will show how this recasts traditional Cornish-Fisher expansions into power series solutions of non-linear ODEs, and give a brief introduction to how this might be applied. 

In the second part of the paper we focus on the issue of building quantiles for the normal distribution based on the recycling approach.  In Section 4 we set out some ideas needed for the development of approximations to quantile functions, however constructed, for any distribution. Section 5 reviews the precision aspects of the current range of Gaussian models in single precision. Section 6 then implements the change of variables approach based on an exponential distribution, and we will then argue that such an approach based on exponential distributions offers better precision characteristics while maintaining class-leading speed. Section 8 makes a corresponding analysis for double precision computation, and additionally introduces the novel idea of generating Gaussian samples by first sampling a Student t distribution with two degrees of freedom. The combination of these two approaches results in a new formula for the Gaussian quantile in double precision that is faster on a GPU than any method known to us, while preserving full double precision on a range more than enough for Monte Carlo work. Section 8 is a brief introduction to managing hyperbolic and Variance Gamma distributions by filtering through an exponential. Section 9 concludes, and relevant codes are supplied in three appendices. 

\section{Quantile mechanics}
If $f(x)$ is the probability density function for a real random variable $X$, the first order quantile ODE is obtained by differentiating Eqn. (1), to obtain:
\begin{equation}
f(w(u))\frac{dw(u)}{du} = 1, \label{fquant}
\end{equation}
where $w(u)$ is the quantile function considered as a function of $u$, with $0 \leq u \leq 1$. Applying the product rule with a further differentiation we obtain:
\begin{equation}
f(w(u))\frac{d^2w(u)}{du^2}+f'(w(u))\biggl(\frac{dw(u)}{du}\biggr)^2 = 0.
\end{equation}
This may be reorganized as
\begin{equation}
\frac{d^2 w(u)}{du^2} = H(w(u)) \left(\frac{dw(u)}{du} \right)^2\ , \label{peaone}
\end{equation}
where
\begin{equation}
H(w) = -\frac{d\ }{dw} \log\{f(w)\}\ .
\end{equation}
and the simple rational form of $H(w)$ for many common distributions, particularly the Pearson family, allows analytical series solutions to be developed \cite{qmone}. This last equation we refer to as the second order quantile equation. 

\subsection{The Recycling ODE}
Now suppose that we make a change of {\it independent} variable in the second order quantile equation Eqn~(\ref{peaone}). We let $v = q(u)$, and regard $w$ as a function of $v$ by writing $w(u) = Q(v)$. Elementary application of the chain rule and some algebra gives us:
\begin{equation}
\frac{d^2 Q(v)}{dv^2}+\frac{q''(u)}{[q'(u)]^2}\frac{dQ(v)}{dv} = H(Q(v)) \left(\frac{dQ(v)}{dv} \right)^2\ , \label{peatwo}
\end{equation}
In general this is a rather awkward differential equation. However, when we regard $q(u)$ as being itself a quantile function, we can make some simplifications. If $q(u)$ is a quantile mapping, it satisfies an ODE of the form
\begin{equation}
\frac{d^2 q(u)}{du^2} = \hat{H}(q(u)) \left(\frac{dq(u)}{du} \right)^2\ , \label{peaone}
\end{equation}
where
\begin{equation}
\hat{H}(w) = -\frac{d\ }{dw} \log\{\hat{f}(w)\}\ .
\end{equation}
and $\hat{f}$ is the probability density function associated with the quantile $q(u)$. So we can simplify the ODE to 
\begin{equation}
\frac{d^2 Q(v)}{dv^2}+\hat{H}(q(u))\frac{dQ(v)}{dv} = H(Q(v)) \left(\frac{dQ(v)}{dv} \right)^2\ , \label{peathree}
\end{equation}
and bearing in mind that $v=q(u)$ we arrive at the ``Recycling Ordinary Differential Equation":
\begin{equation}
\frac{d^2 Q(v)}{dv^2}+\hat{H}(v)\frac{dQ(v)}{dv} = H(Q(v)) \left(\frac{dQ(v)}{dv} \right)^2\ , \label{peafour}
\end{equation}
We now turn to two particularly interesting cases, rather inspired by Mikosch's suggestions \cite{mikosch}.
\subsection{The Recycling ODE for a Gaussian background}
In this case we have the following obvious sequence of manipulations:
\begin{equation}
\hat{f}(x) = \frac{1}{\sqrt{2\pi}}e^{-x^2/2}
\end{equation}
\begin{equation}
\log\hat{f}(x) = -1/2 \log(2 \pi)- x^2/2
\end{equation}
\begin{equation}
\frac{d\ }{dx}\log\hat{f}(x) = -x
\end{equation}
\begin{equation}
\hat{H}(v) = v
\end{equation}
and we arrive at the Recycling ODE for a Gaussian background as 
\begin{equation}
\frac{d^2 Q(v)}{dv^2}+v \frac{dQ(v)}{dv} = H(Q(v)) \left(\frac{dQ(v)}{dv} \right)^2\ , \label{peatwo}
\end{equation}
This is an interesting example to consider for target distributions along the entire real line. 

\subsection{The Recycling ODE for a one-sided exponential  background}
In this case we have the following obvious sequence of manipulations:
\begin{equation}
\hat{f}(x) = e^{-x},\ \ 
\log\hat{f}(x) = - x,\ \  \frac{d\ }{dx}\log\hat{f}(x) = -1,\ \ \hat{H}(v) = 1
\end{equation}
and we arrive at the Recycling ODE for a exponential background as 
\begin{equation}
\frac{d^2 Q(v)}{dv^2}+\frac{dQ(v)}{dv} = H(Q(v)) \left(\frac{dQ(v)}{dv} \right)^2\ , \label{peatwo}
\end{equation}
This is an interesting example to consider for target distributions along the positive real line. For distributions that are asymptotically exponential in both directions it can be used in two pieces.

\section{Example with a Gaussian background}
In a Gaussian background we work with the Recycling ODE in the form
\begin{equation}
Q''+v\,Q'= H(Q) (Q')^2
\end{equation}
where the explicit dependence on $v$ is suppressed for brevity, and ' denotes $d/dv$. The target distribution is encoded through the function $H$. Note that it is not required in any sense that the target distribution is ``close'' to, or asymptotic to a Gaussian. This is an {\it exact} relationship governing the function $Q$ that is the composition of the Gaussian CDF followed by the quantile of the target distribution. But such a relationship must contain all information relevant to the creation of an expansion of one distribution in terms of another. In particular, we should be able to re-create known and new expansions of Cornish-Fisher type. Generalized Cornish-Fisher expansions have been considered in the notable paper by Hill and Davis \cite{hilldavis}, but the step to considering the matter as the solution of a single differential equation is, so far as this author is aware, a new one. 

\subsection{The Student distribution}
This is an interesting case for several reasons:
\begin{enumerate}
\item We can illustrate the method;
\item We can recover a well known asymptotic series;
\item We can develop that series to arbitrary numbers of terms;
\item We can explore the limitations of the known series;
\item We can develop an alternative numerical method and explore purely numerical options.
\end{enumerate}
The density and associated $H$-function for the Student case can be written down as
\begin{equation}
f_{T_{n}}(x) =     \frac{1}{\sqrt{n\pi}} \frac{\Gamma[(n+1)/2]}{\Gamma[n/2]}\biggl(1+ \frac{x^2}{n} \biggr)^{-\frac{n+1}{2}}
            \ ,\ \ \ H_{T_{n}}(Q) = \biggl(1 + \frac{1}{n}\biggr) \frac{Q}{1+Q^2/n}
\end{equation}
where $n$ is the degrees of freedom (not necessarily an integer), and the Recycling ODE can be written in the form
\begin{equation}
\biggl(1+ \frac{Q^2}{n}\biggr)\biggl(Q''+v\, Q'\biggr) =  \biggl(1+ \frac{1}{n}\biggr)Q (Q')^2\ \end{equation}
We note that if we let $n \rightarrow \infty$ we obtain
\begin{equation}
Q''+v\,Q' = Q (Q')^2
\end{equation}
and this has the desired solution $Q=v$. More generally we can look at series solutions, but should be mindful of the fact that the term $Q^2/n$ is present - this is a hint that the behaviour of series for $Q \ll \sqrt{n}$ and $Q\gg\sqrt{n}$ could be rather different. Such considerations do not always apply if one is thinking in a purely asymptotic framework. For any {\it finite} $n$, no matter how large, there will always be values of $Q$ such that the behaviour is far from Gaussian. This was alluded to in \cite{shawjcf06}, where it was noted that the known Cornish-Fisher expansion always goes wrong in the tails at some point.

We also need to consider boundary conditions. The derivative of any ordinary quantile function at a point is the inverse of the PDF at the corresponding quantile. We first work around the point $u=1/2$ which corresponds to $v=0$ in the Gaussian coordinate. If $z(u)$ and $t(u)$ are the ordinary quantiles for the Gaussian and Student distributions, then we have
\begin{equation}
\begin{split}
z(1/2) &= 0,\\
z'(1/2) &= \sqrt{2\pi}\\
t(1/2) &= 0,\\
t'(1/2) &= \sqrt{n\pi} \frac{\Gamma[n/2]}{\Gamma[(n+1)/2]}
\end{split}
\end{equation}
It follows that the centre conditions we wish to apply to the Recycling ODE are just:
\begin{equation}
\begin{split}
Q(0) &= 0,\\
Q'(0) &= \gamma \equiv \sqrt{\frac{n}{2}} \frac{\Gamma[n/2]}{\Gamma[(n+1)/2]}
\end{split}
\end{equation}
where the latter expression $\gamma$ arises as the ratio of the derivatives.

\subsection{The central expansion}

We now develop a series solution about the centre, and we expect that it will be reasonable to treat the solution as ``close to Gaussian'' if $Q^2 \ll n$. We assume, as both the normal and Student quantiles are symmetric, that
\begin{equation}
Q(v) \sim \sum_{k=0}^{\infty}c_k v^{2k+1}
\end{equation}
where $c_0 = \gamma$. We use the tilde notation to indicate that at this point we have no presumption as to whether the resulting series will be convergent for all $v$ or form some kind of asymptotic series. We find that
\begin{equation}
\begin{split}
c_1 &= \frac{(n+1) \gamma ^3-n \gamma }{6 n} \\
c_2 &= \frac{\left(7 n^2+8 n+1\right) \gamma ^5+\left(-10 n^2-10 n\right) \gamma ^3+3 n^2
   \gamma }{120 n^2}
\end{split}
\end{equation}
Subsequent terms may be generated by iteration of the RODE, and in this case, after some algebra, we find that
\begin{equation}
\begin{split}
(2i+3)(2i+2)c_{i+1} =& -(2i+1) c_i \\
&+ \sum_{l=0}^i \sum_{m=0}^{i-l}a_{lm}(n)c_{i-l-m}c_l c_m \\
&- \frac{\theta(i)}{n}\sum_{l=0}^{i-1}\sum_{m=0}^{i-1-l}(2m+1) c_{i-1-l-m}c_lc_m,
\end{split}
\end{equation}
where $\theta(0)=0, \theta(i)=1$ if $i \geq 1$, and
\begin{equation}
a_{lm}(n) = (1+\frac{1}{n})(2l+1)(2m+1)-\frac{2}{n}m(2m+1)
\end{equation}

\subsection{The tail expansion}

We now develop a series solution about the right tail $Q \rightarrow \infty$. We begin by assuming that  $Q^2 \gg n$. The Recycling ODE becomes
\begin{equation}
Q(Q''+v Q') =  (n+1) (Q')^2
\end{equation}
Following some experimentation, we make the change of variables
\begin{equation}
P(v) = \frac{1}{Q(v)^n}
\end{equation}
and this reduces the ODE to
\begin{equation}
P''(v) + v P'(v) = 0
\end{equation}
The solution of this satisfying the condition that $P(v) \rightarrow 0$ as $v \rightarrow \infty$ is
\begin{equation}
P(v) \propto {\rm erfc} (\frac{v}{\sqrt{2}})
\end{equation}
and we deduce that for some constant $c$,
\begin{equation}
Q(v)  \sim c \biggl[\frac{1}{2}{\rm erfc} (\frac{v}{\sqrt{2}})\biggr]^{-1/n}
\end{equation}
We see that the solution has emerged naturally as
\begin{equation}
Q(v)  \sim c \biggl[1-\Phi(v)\biggr]^{-1/n}
\end{equation}
where $\Phi$ is the Gaussian CDF. The asymptotic differential equation is scale invariant so we have to determine $c$ by other means. It is possible that it might be possible to determine it by a matching argument, but it is simpler to now appeal to other known properties of the Student distribution. In \cite{shawjcf06} the tail behaviour of the Student CDF was determined (see Eqns. (60-62) of \cite{shawjcf06}) and we can deduce that
\begin{equation}
c = \sqrt{n} \biggl[n \sqrt{\pi} \frac{\Gamma(n/2)}{\Gamma((n+1)/2)}  \biggr]^{-1/n}
\end{equation}
If we step back from these calculations it becomes clear what is happening. The Recycling ODE is starting to reconstruct a solution that combines the change of variable $w = 1-\Phi(v)$ with the asymptotic power series of the ordinary Student quantile. 

\subsection{Comparison with traditional asymptotics}
Expansions of Cornish-Fisher type can be found in the statistics literature. One that is reasonably well known is the expansion of the Student random variable $t$ in terms of the Gaussian random variable $z$, for larger values of the degrees of freedom $n$. 
It is quoted, for example, as identity 26.7.5 of \cite{amsteg}. 
\begin{equation}
\begin{split}
t =& z+\frac{z^3+z}{4 n}+\frac{5 z^5+16 z^3+3 z}{96 n^2}+\frac{3 z^7+19 z^5+17 z^3-15 z}{384 n^3}\\
&+\frac{79 z^9+776 z^7+1482 z^5-1920 z^3-945 z}{92160 n^4}+\dots
\end{split}\label{cfstu}
\end{equation}
An equation of true Cornish-Fisher type (cf identity 26.2.49 of \cite{amsteg}) can be obtained by transforming (provided $n>2$) to a variable $s$ with {\it unit variance}: $s = t\sqrt{1-2/n}$ and re-expanding in inverse powers of $n$. That Eqn.~(\ref{cfstu}) is somehow incomplete is evident by the fact that $z$ appears in every term, $z^3$ in all but the first, and so on. The matter is resolved nicely by first observing that
\begin{equation}
\gamma = 1+\frac{1}{4 n}+\frac{1}{32} \left(\frac{1}{n}\right)^2-\frac{5}{128}
   \left(\frac{1}{n}\right)^3-\frac{21
   \left(\frac{1}{n}\right)^4}{2048}+\frac{399
   \left(\frac{1}{n}\right)^5}{8192}+O\left(\left(\frac{1}{n}\right)^6\right)\ ,
\end{equation}
which sums up all the $z$-terms. Similarly
\begin{equation}
c_2 = \frac{1}{4 n}+\frac{1}{6} \left(\frac{1}{n}\right)^2+\frac{17}{384}
   \left(\frac{1}{n}\right)^3-\frac{1}{48}
   \left(\frac{1}{n}\right)^4-\frac{17
   \left(\frac{1}{n}\right)^5}{8192}+O\left(\left(\frac{1}{n}\right)^6\right)
\end{equation}
and so on. So the series solution of the differential equation constitutes a re-summation of the known asymptotic series where the coefficient of each power of $z$ is computed exactly.  This means that we can use the series without assuming that $n$ is large, as it is a series valid in some domain for $z$, the existence of which does not depend on $n$ being large.

\subsection{Accuracy and numerical methods}
We now turn to the quality of the results. This can be assessed precisely by the use of an exact representation of the composite function $F_N^{-1}(\Phi(z))$, where $\Phi$ is the normal CDF and $F_n$ the Student CDF. The exact formula for the Student CDF for all real $n$ is given in terms of inverse beta functions by Shaw \cite{shawjcf06}, and there are known simpler forms for $n=1,2,4$. These are also given in \cite{shawjcf06} and are also now available on the {\it Wikipedia} page on quantile functions \cite{wikiquantile}. The case $n=4$ is an interesting case as it is known exactly, in the boundary case where kurtosis is infinite, and there is some evidence from work by Fergusson and Platen \cite{platen} that it is a good case for modelling daily world index log-returns. We shall therefore develop this in some detail. It turns out that working as far as $c_{10}$ is a useful point. A detailed calculation shows that the precision (i.e. relative error) of the central power series is then less than $2 \times 10^{-5}$ on $|z| < 4$. Given that $z$ is a standard Gaussian this is within the range of $\pm 4$ standard deviations for the underlying Gaussian. 

For this case we find that 
\begin{equation}
\gamma = \frac{4}{3}\sqrt{\frac{2}{\pi}} \sim 1.06384608107048714 
\end{equation}
and the full C-code form for the central series is, with $y=z*z$,
\begin{verbatim}
t = z*(1.06384608107048714 + 
      y*(0.0735313753642658509 + 
        y*(0.00408737916150927847 + 
           y*(0.000157376276663230562 + 
              y*(4.31939824140363509e-6 + 
                 y*(9.56881464639227278e-8 + 
                    y*(2.09256881803614446e-9 + 
                    y*(3.87962938209093352e-11 + 
                    y*(2.72326084541915671e-13 + 
                    (2.90528930162373328e-15 + 
                    4.59490133995901375e-16*y)*y)))))))
        ))
        \end{verbatim}
 To treat the tail regions $|z|>4$ with corresponding accuracy when $n=4$ it is sufficient to use just {\it two} terms of the known tail series. This gives us, in general, for the positive tail (the negative tail being treated by symmetry)
 \begin{equation}
 \begin{split}
 w &=(1-\Phi(z))n\sqrt{\pi}\frac{\Gamma(n/2)}{\Gamma((n+1)/2)} \\
 t &= \sqrt{n} w^{-1/n}(1 -\frac{n+1}{2(n+2)} w^{2/n})
 \end{split}
 \end{equation}
 and for the case $n=4$:
  \begin{equation}
 \begin{split}
 w &=(1-\Phi(z))\frac{16}{3}\\
 t &= 2 w^{-1/4}(1 -\frac{5}{12} w^{1/2})
 \end{split}
 \end{equation}
 The optimal crossover is then in fact at $z=3.93473$ with maximum relative error less than $1.4 \times 10^{-5}$ over the entire range of $z$
\subsection{Purely numerical methods}
The analysis for the Student t case, although rather specialized, also allows the appraisal of direct numerical schemes. The direct numerical solution of the RODE can be done using standard methods. Within {\it Mathematica} version 6, the use of {\tt NDSolve} with high precision and accuracy goals, explicit Runge-Kutta and sixth-order differences leads to an precision of better than $5 \times 10^{-8}$ on the range $|z| < 6$, which is excellent. Of course, one must also consider sampling efficiency issues arising from such interpolated numerical schemes, but they can be made the basis of a further, e.g. rational approximation if speed is an issue. Such a numerical scheme will be exploited in the examples considered below. There is clearly much more scope for transforming between Student distributions of different $n$ and Gaussians, and later in this paper we will provided an example of transforming a $T_2$ to a Gaussian.  Given that there closed-forms for the quantiles for $T_1, T_2$ and $T_4$ these might serve as useful intermediate distributions for other applications as well, but this will be explored elsewhere.

\section{Issues in building a better quantiles}
In this section we consider the issues involved in building more accurate and/or faster quantiles for live use on computer simulations. These will be important for {\it any} distribution, but the particular focus here will be on finding new forms of quantile for the normal distribution. The construction of the normal quantile, also know as ``probit'', has a long and interesting history - see \cite{qmone} and the references therein for details . This is important given its base case use in general modelling applications, and in particular given its massively repeated use in the solution of stochastic differential equations. The issues we consider are
\begin{enumerate}
\item The target precision;
\item The domain over which the target is to be rigorously achieved;
\item The behaviour outside the target domain;
\item The choice of intermediate probability distribution to be used in the recyling scheme;
\item The algorithm's performance on typical CPU and GPU architectures.
\end{enumerate}
The last point applies to the particular approach developed here, but the first four points are relevant to any approach. In the following sections we explore these issues and the associated choices. This gives some basic parameters relevant to our general scheme of implementing ``Quantile Mechanics''. In the past, diverse choices have been made for precision and domain, and we think it appropriate to better characterize these in mathematical form.

\subsection{The precision target}
Given an exact representation of a function $Q(u)$ and an approximation of it $Q_a(u)$ we can consider various norms on their differences. We will work in the supremum norm for the relative precision over a domain ${\cal D}$ to be specified, and seek to minimize the relative error
\begin{equation}
\epsilon_R = \sup_{\cal D}\bigg\lvert {\frac{Q_a(u)}{Q(u)}-1}\bigg\rvert\ .
\end{equation}
What this seeks to do is to maximize the quality of the significant figures of the result. We do have to bear in mind that such a goal may be interpreted in two ways. The first approach, that will be taken here, is to set such a target for the theoretical precision of an approximation (typically polynomial or rational), given an abundance of computer precision. The second approach is to set the target based on the actual errors in a given limited precision environment. While the second approach is clearly more pragmatic, the answers we will get will depend significantly on implementation/hardware/compiler details. We will therefore consider the first approach, which allows us to develop the analysis mathematically, and which is in common use in modelling. So, for example, the first form (i.e. the simple form without an application of the Newton-Raphson-Halley iteration) of the normal quantile developed by Acklam \cite{acklam} sets
\begin{equation}
\epsilon_R \lesssim 1.15 \times 10^{-9}
\end{equation}
and our own computational analysis of Wichura's AS241 \cite{wichura} suggests that in that case
\begin{equation}
\epsilon_R <  7.45 \times 10^{-17}
\end{equation}
when worked out in high precision arithmetic. The Acklam error bound is also attained in double-precision arithmetic, while AS241 errors (in our implementation) peak at about $5 \times 10^{-16}$.

In our own studies we have chosen the set the target theoretical precision based on IEEE levels\footnote{We are grateful to M. Pont of NAG for advice on these points}. Given that  IEEE single
precision has a 24 bit mantissa,  we will aim in single, or {\it float} precision for
\begin{equation}
\epsilon_{Rf} =  2^{-25} = 2.98 \times 10^{-8}\ .
\end{equation}
Since IEEE double precision has 53 bits of precision we will set
\begin{equation}
\epsilon_{Rd} =   2^{-54} = 5.55 \times 10^{-17}
\end{equation}
for {\it double} precision. So these values are half the values often quoted as ``machine epsilon'' for the two cases. The matter is somewhat ethereal in view of different possible rounding methods and compiler variations, and in some circumstances a float target of $2^{-23} = 1.19 \times 10^{-7}$ may be perfectly acceptable. The error target might also be expressed in a different norm. So, for example, M. Giles  works with an $L_2$ on the relative error and a theoretical target of about $10^{-7}$ in his GPU Computing Gems article \cite{newgiles}. That work also includes a very useful discussion of realized errors.

\subsection{The target domain}
This is a somewhat subtle matter and massively different choices are possible. The quantile mapping for distributions along the entire real line is from the domain $(0,1)$ to the range that is the real line. In practice computation will take place on an interval
\begin{equation}
U(\epsilon_L, \epsilon_R) = (\epsilon_L, 1-\epsilon_R)
\end{equation}
and the question for this section is the choice of $\epsilon_L, \epsilon_R$. The difficulty is that there is some asymmetry in the matter, as is well discussed on P. Acklam's web page \cite{acklam}. It is clear that the value of $\epsilon_R$ should be related to the machine epsilon, as there is simply no possibility of getting closer to unity in finite precision arithmetic. For example, Acklam's discussion considers the errors up to both the points $1-2\epsilon_{Rd}$ and $1-\epsilon_{Rd}$. 

The problems are more to do with the choices made for $\epsilon_L$. One might argue on symmetry grounds that one should take $\epsilon_{L} = \epsilon_{R}$. Indeed, if the analysis is done for the inverse error function, rather than the quantile, this is entirely natural, and is the approach taken by Giles \cite{newgiles}, with a symmetric range for {\it erfinv(x)} from $(-1+\epsilon_G, 1-\epsilon_G)$ where $\epsilon_G$ is about $5.6 \times 10^{-8}$ for float and $1.16 \times 10^{-16}$ for double, based on the largest required values for $w = -\log(1-x^2)$ of $16$ and $36$. 

While such symmetric choices are sensible for {\it erfinv}, it seems to us that they do not go far enough for the case of the quantile function with the domain of the unit interval, with lower bound at zero. Acklam \cite{acklam} takes the view that the quantile needs to have a controlled precision down to at least $2.23 \times 10^{-308}$, which is the smallest full precision number that  can be given in double precision, and his results show controlled errors on a still larger range. 
 
If one is interested in representing a mathematical function for diverse applications to as much precision as possible, we take the view that a domain on the lines of the extended form proposed by Acklam is entirely appropriate. However, for Monte Carlo applications this is massively over-engineered, but we argue also that the Giles method does not quite go far enough. To make the case for a different choice, we consider the typical scope of a Monte Carlo simulation. 

Consider a set of $N$  independent uniformly distributed random numbers, $\{U_i \}, i=1,\dots,N$ taken from the interval $(0,1)$, and let 
\begin{equation}
U_{\rm min} = \min  \{U_i \}\ .
\end{equation}
The distribution of $U_{\rm min} $ is easily worked out using elementary probability:
\begin{equation}
P(U_{\rm min}  < u) = 1 - P(U_{\rm min}  \geq u) = 1 - P(U_i  \geq u,\  \forall i) = 1 - P(U_i  \geq u)^N
\end{equation}
and given that each $U_i$ has a uniform distribution we see that 
\begin{equation}
P(U_{\rm min}  < u) =1 - (1-u)^N 
\end{equation}
and it is then easily computed that
\begin{equation}
E(U_{\rm min}) =\frac{1}{N+1} \ ,\ \ {\rm Var}(U_{\rm min})  = \frac{N}{(N+1)^2(N+2)}\ ,
\end{equation}
and we see that the average smallest value is about $1/N$ for large $N$, but the standard deviation is about the same as the mean. 
Inverting the distribution function we see that the quantile function for this minimum is given by
\begin{equation}
Q_{\rm min} (v) =   1 - (1-v)^{1/N} \sim -\frac{1}{N} \log(1-u) \ {\rm as}\  N \rightarrow \infty
\end{equation}
which is an exponential distribution. For large $N$ and small $U$ this quantile function is approximately
\begin{equation}
Q_{\rm min} (v) \sim \frac{u}{N} \ .
\end{equation}
In other words, we have about a $1\%$ chance of encountering numbers lower than 
\begin{equation}
\frac{1}{100*N}
\end{equation}
in our simulation. Even a rather modest Monte Carlo simulation therefore has a significant chance of producing numbers going well below the machine epsilon in float mode, though not in double precision. 

One might argue that the influence of a few such small numbers on the overall computation of expectations might be small. The difficulty with such a view is that one can imagine situations where the computation is dominated by minima (lookbacks, options on the worst performer of a basket, systems coupled by a Clayton copula with a lot of downside correlation, deeply out of the money Puts, and so on). Furthermore, the use of antithetic methods might result in low values being reflected to high values as well after the application of the target distribution's quantile function. 

In view of this we argue that the largest allowable low-end truncation point should be no higher than about $10^{-10}$, meaning that $99\%$ of  random numbers would be mapped to the target distribution accurately for samples up to $N=10^8$ in size. Options should be provided to go to smaller values for special problems requiring very large samples. What this means is that in the left tail the machine epsilon bound will be ignored in float mode, and replaced by a number no higher than $10^{-10}$. Bearing in mind that our algorithms will be symmetric in any case, we will take our target region as
\begin{equation}
(\varepsilon, 1- \varepsilon)
\end{equation}
where
\begin{equation}
\varepsilon < 10^{-10} \ {\rm for\  {\it float}\ and \ }\ 
\varepsilon < 5.55 \times 10^{-17} \ {\rm for\  {\it double}}\ \ .
\end{equation}
In the case of double precision, it seems to us that the machine epsilon level is already small enough to manage a super-large Monte Carlo simulation. We see no need to go down to the smallest full precision float, about $10^{-38}$, let alone to the smallest possible double, for Monte Carlo applications, as such small numbers are very unlikely to occur.  

\subsection{Out of domain sample control }
Having made the choices in the previous sub-section, we wish to insure that the values of random variables generated if $u < \varepsilon$ are not too poor in their error characteristics. That is, we require that the error growth below this level is reasonable. For example, if the relative error in a one in a billion sample point is perhaps $10^{-6}$, it will not matter, but much larger relative errors cannot be accepted. 

\subsection{Choice of filtering distribution}
Moving now to the particular case of the normal quantile, we have several options for intermediate distributions of interest. The choices we will explore are suitable scalings of:
\begin{itemize}
\item The one-sided exponential distribution;
\item The Student t distribution with two degrees of freedom;
\item An ``almost-normal'' distribution;
\item An ``almost-chi-squared'' distribution. 
\end{itemize}
The one-sided exponential distribution has PDF, CDF and Quantile given by, for $x>0$, $0<u<1$, 
\begin{equation}
f_E(x)  = e^{-x}\ , F_E(x) = 1-e^{-x}\ , Q_E(u) = -\log(1-u)\ .
\end{equation}
The Student t distribution with two degrees of freedom has PDF, CDF and Quantile given by, for all real $x$, $0<u<1$, 
\begin{equation}
f_t(x)  = \frac{1}{(2+x^2)^{3/2}}\ , F_t(x) = \frac{1}{2} + \frac{x}{2\sqrt{x^2+2}}\ , Q_t(u) = \frac{2u-1}{\sqrt{2u(1-u)}}\ .
\end{equation}
The almost-normal distribution\footnote{This is a distribution one of us (WS) invented for an examination question, and it is left to the reader to compare its density, distribution and quantile with the exact normal.} is literally one that is very close to a standard Gaussian but has a simple closed-form quantile.  The density and distribution function are given by
 \begin{equation}f_{AN}(x) = \frac{e^{-\frac{2 x^2}{\pi }} |x|}{\pi  \sqrt{1-e^{-\frac{2 x^2}{\pi }}}}\end{equation}
 \begin{equation} F_{AN}(x) = \frac{1}{2}+ \frac{1}{2} \text{sign}(x) \sqrt{1-e^{-\frac{2 x^2}{\pi }}} \end{equation}
and it is then an elementary exercise to compute the quantile function as
\begin{equation}
Q_{AN}(u)  =  {\rm sign}(u-1/2)*\sqrt{-\frac{\pi}{2} \log(4u(1-u))} \ .
\end{equation}
The almost-chi-squared distribution is that of a random variable that is the square of an almost-normal distribution. The density, distribution and quantile are given by
\begin{equation}f_{AC}(x) = \frac{e^{-\frac{2 x}{\pi }} }{\pi  \sqrt{1-e^{-\frac{2 x}{\pi }}}}\end{equation}
 \begin{equation} F_{AC}(x) =  \sqrt{1-e^{-\frac{2 x}{\pi }}} \end{equation}
\begin{equation}
Q_{AC}(u)  = -\frac{\pi}{2} \log(1-u^2) \ .
\end{equation}
This distribution can also be simulated via the $2-1$ mapping $u\rightarrow Q_{AN}^2(u)$, that has the same image as $ Q_{AC}(u)$, 
\begin{equation}
Q_{AN}^2(u)  = -\frac{\pi}{2} \log(4u(1-u)) \ .
\end{equation}
It is informative to look at the current list of approximate normal quantiles to see how prevalent these changes of variable are.
\begin{itemize}
\item Our earlier branchless formulae \cite{sbarxiv} relied on the one-sided exponential distribution;
\item Giles' method \cite{newgiles} is essentially based on transforming the almost-gamma in the central region and the almost-normal in the tail;
\item Acklam's non-iterated formula \cite{acklam} is the asymptotic form of almost-normal in the tail, as is AS241 \cite{wichura}. 
\end{itemize}
The Moro method \cite{moro} uses a double-log in the tail region. 

\subsection{Computing architecture and compiler issues}
In this section we consider some of the details of computer science that impact the mathematical analysis in the context of trying to be as efficient as possible.
\subsection{Branching management}
On a CPU there is normally very sophisticated management of algorithm branching, to the extent that (a) having a branch at all is managed efficiently and (b) a slow branch will not compromise the operation of a fast branch. On a GPU we have to confront the issue of {\it warp divergence}. A warp is a group of 32 threads (in current generation NVIDIA GPUs) that evaluate concurrently. If there is an IF statement and one thread needs to execute a different branch, then the current behaviour will be for all threads to execute both branches. This is especially problematic in quantile evaluation where typically 2 or more regions are considered for separate evaluation, with (as noted above) a more computationally expensive change of variables in the tail. There are two existing approaches:
\begin{itemize}
\item create an essentially branchless algorithm where one formula applies for all relevant values;
\item lower the probability of a warp divergence to a much smaller value.
\end{itemize} 
The first method is that taken by an earlier version of this study \cite{sbarxiv}, and the second approach is that taken by Giles \cite{newgiles}. The essential question here is on the balance of effort required to go essentially branchless versus that where an unlikely branch remains but with a simpler formula in the middle region. In our final model for double precision we will develop a third approach based on the idea of warp voting employing the $all()$ function, in which one of two essentially branchless algorithms will be chosen depending on the warp as a whole.

\subsubsection{FMA support}
The presence in some architectures of a ``fused multiply-add'' (FMA) operation in a single hardware operation creates a performance bias in support of algorithms that rely on repeated applications of operations such as are found in Horner's scheme for fast polynomial evaluation, where, e.g. a general cubic would be evaluated as
\begin{equation}
p \rightarrow d*x + c\ ,\ \ p \rightarrow p*x + b\ ,\ \ p \rightarrow p*x + a\ .
\end{equation}
Given that CUDA supports FMA we have found that there is no benefit in performing any form of pre-processing to work out polynomials or rational approximations using other devices to reduce the number of multiplication operations. So, for example, the method of quadratic factorization (see e.g. section 7 of \cite{rabin}) has been found to produce no speed-up despite reducing the number of multiplications, and special methods available for polynomials of specific degree (e.g. a trick for degree 6 described in exercise 6 for Section 7.2 of  \cite{rabin}), both produce no speed-up but also generate a loss of precision due to several subtractions. So will will work with iterated linear maps on the basis of FMA support whenever a polynomial or rational function has to be evaluated, though we should note that in non-FMA architectures that quadratic factorization may be a useful route for further optimization. 

\subsubsection{Optimization of simple function evaluation}
Given the comments made in Section 4.4 it is important to understand the costs of evaluating {\it log} and {\it sqrt} compared to working out polynomials of various size. In float mode both log and sqrt are similarly expensive, so that for the main block of the algorithm for the central quantile, one application of one of these functions is helpful, but applying both is inefficient. For example, the main brake on Acklam's formula is the slow computation of the tail, which as we have noted involves both a  {\it log} and {\it sqrt} application. 

One of the surprises of our analysis is the benefit of the efficiency in CUDA of the reciprocal square root function $rsqrt$, especially in double precision. This means that the $T_2$ distribution, suitable scaled, is an unexpectedly efficient filtering distribution. We will return to this later. 

\section{The current Gaussian float quantile zoo considered on precision}
In this section we will explore the precision characteristics of the existing range of commonly employed methods, in both {\it float} and {\it double} precision. For {\it float} mode we will consider
\begin{itemize}
\item Moro's modification of the Beasley-Springer method \cite{moro};
\item Acklam's non-iterated method \cite{acklam};
\item Our earlier constructions of an essentially branchless formula for GPUs\cite{sbarxiv};
\item Giles' construction of a highly-efficient formula for GPUs \cite{newgiles};
\end{itemize}
In each case we explore the theoretical precision of the method, using an implementation in Mathematica\footnote{The notebooks employed are available on request from the corresponding author} employing high-precision arithmetic. In this way we see the essential properties of the formula employed. We consider the precision both inside the region certified by the originators for precision, and, where appropriate, the behaviour outside. First, however, we explore the overall relative precision in the unit interval. This will show the precision for the majority of samples, and is also indicative of the certified theoretical precision. 
\begin{figure}[hbt]
\centering
\includegraphics[scale=0.6]{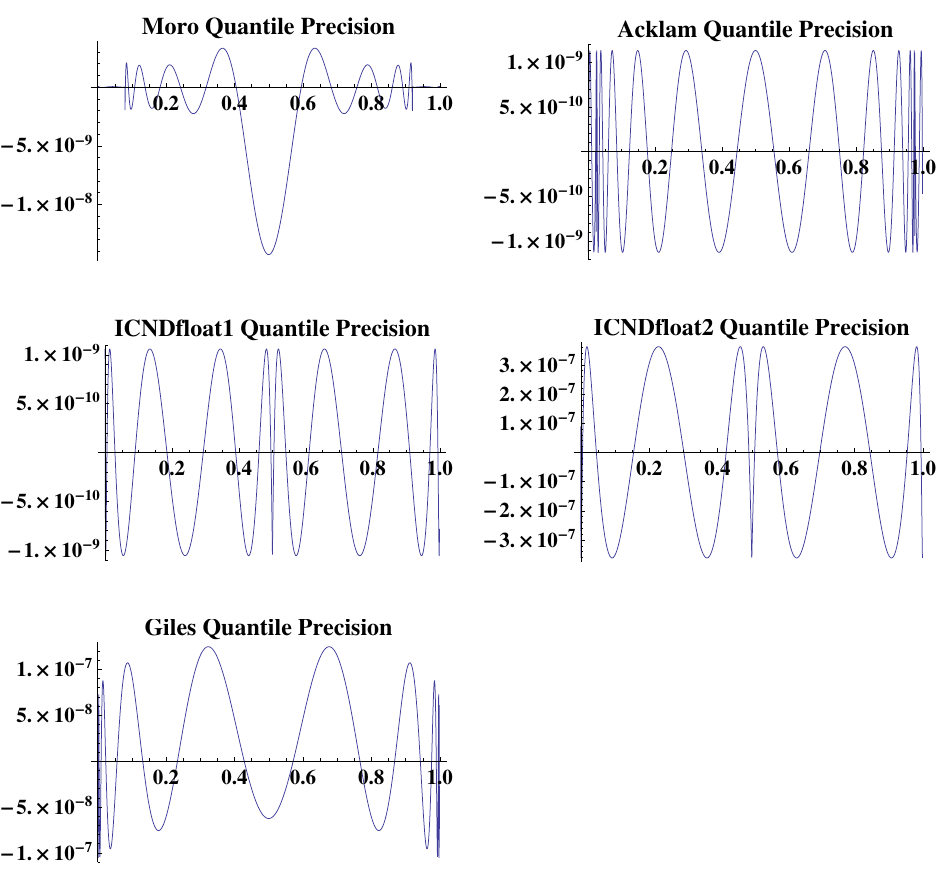}\label{oldprec}
\caption{Theoretical relative precision of standard quantiles - main region}
\end{figure}
In interpreting Fig. 1, we should note that the Moro formula was certified and constructed based on absolute rather than relative error, and that Giles' formula is based on a weighted $L_2$ model rather than a supremum norm.  Next we consider the behaviour in the left tail region, as show in Fig. 2.
\begin{figure}[hbt]
\centering
\includegraphics[scale=0.6]{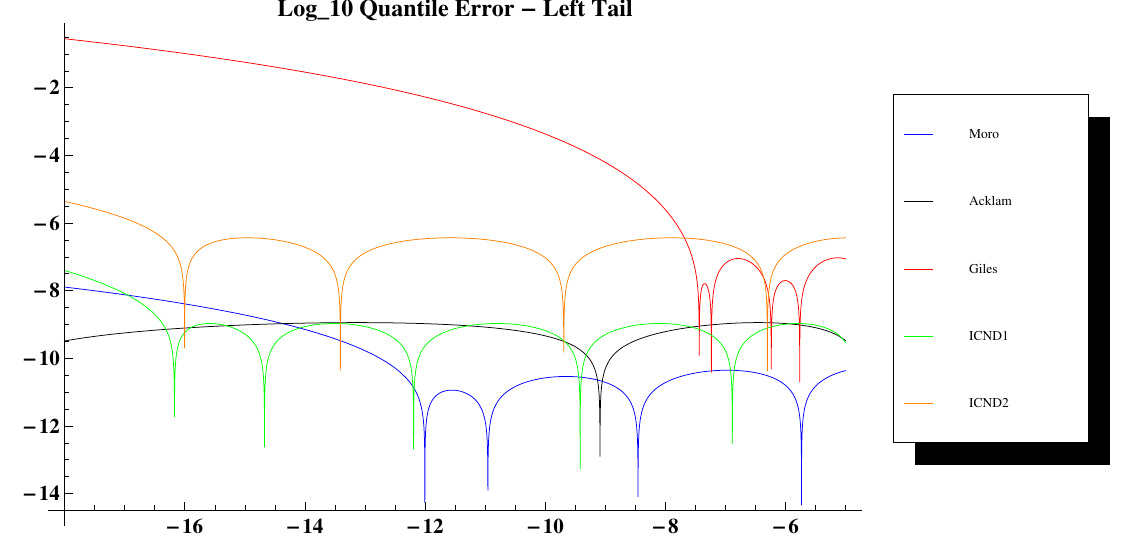}\label{lefttail}
\caption{Theoretical relative precision of standard quantiles in left tail}
\end{figure}
In Fig. 2 we have plotted the base-10 logarithm of the absolute value of the precision as a function of the base-10 log of the input $u$.  We should recall that the Giles formula was not designed to perform well for points closer to zero than machine epsilon. The growth in the error before we get down to that level is, however, a concern. We have shown our two early branchless approximations ICND1 and ICND2, where ICND1 \cite{sbarxiv} was designed to have similar precision to the Acklam method, and ICND2 was a quicker less precise option. 

These plots suggest that there is some room for improvement, at least as regards the criteria given in Sections 4.1 and 4.2. We shall now proceed to consider the construction of quantiles optimized for speed subject to a precision goal of $
\epsilon_{Rf} =  2^{-25} = 2.98 \times 10^{-8}$ working down to below $10^{-10}$, based on the essentially branchless approach and an exponential filtering distribution.

\section{Normal samples from exponential}
Now we consider the construction of normal samples from exponential samples, and proceed to a detailed practical implementation. We work on the right hand region and extend the mapping to the left region by odd symmetry. The recyling ordinary differential equation in the right hand region, $v\geq 0$ is simply
\begin{equation}
\frac{d^2 Q}{dv^2} + \frac{dQ}{dv} = Q \biggl(\frac{dQ}{dv} \biggr)^2
\end{equation}
with the initial conditions $Q(0)=0$, $Q'(0) = \sqrt{\pi/2}$. This has the formal solution
\begin{equation}
Q(v) = \Phi^{-1}(1-1/2 e^{-v})
\end{equation}
where $\Phi$ is the normal CDF. To extract useful representations we proceed as follows.
This equation may first be solved by the method of series. However, the resulting solution turns out to be an asymptotic series best used to a small number of terms in a neighbourhood of $v=0$. The series solution is easily found to be, using exact coefficients:
\begin{equation}
\begin{split}
Q(v) =& \sqrt{\frac{\pi }{2}} v-\frac{1}{2} \sqrt{\frac{\pi }{2}}
   v^2+\frac{\left(2 \sqrt{\pi }+\pi ^{3/2}\right) v^3}{12
   \sqrt{2}}-\frac{\left(\sqrt{\pi }+3 \pi ^{3/2}\right) v^4}{24
   \sqrt{2}}\\
   &+\frac{\left(4 \sqrt{\pi }+50 \pi ^{3/2}+7 \pi
   ^{5/2}\right) v^5}{480 \sqrt{2}}
   -\frac{\left(4 \sqrt{\pi }+180
   \pi ^{3/2}+105 \pi ^{5/2}\right) v^6}{2880
   \sqrt{2}}\\
   &+\frac{\left(8 \sqrt{\pi }+1204 \pi ^{3/2}+1960 \pi
   ^{5/2}+127 \pi ^{7/2}\right) v^7}{40320 \sqrt{2}}\\
   &-\frac{\left(2
   \sqrt{\pi }+966 \pi ^{3/2}+3675 \pi ^{5/2}+889 \pi ^{7/2}\right)
   v^8}{80640 \sqrt{2}}\\
   &+\frac{\left(16 \sqrt{\pi }+24200 \pi
   ^{3/2}+194628 \pi ^{5/2}+117348 \pi ^{7/2}+4369 \pi ^{9/2}\right)
   v^9}{5806080 \sqrt{2}}\\
   &-\frac{\left(16 \sqrt{\pi }+74640 \pi
   ^{3/2}+1190700 \pi ^{5/2}+1493520 \pi ^{7/2}+196605 \pi
   ^{9/2}\right) v^{10}}{58060800 \sqrt{2}}\\
   &+O\left(v^{11}\right)
\end{split}
\end{equation}
While this expression is interesting, it does not work far enough out to be of much practical use, so a different approach is needed - if we wish to retain the use of the above expression we would need to patch in another algorithm. One could consider solving the differential equations about several points. However, as already noted, important point for modern computation is to try to avoid ``IF'' statements in the computer implementation. Such branches do not make use of the best features of modern GPU systems, such as the NVIDIA Tesla system \cite{nvidia}. The standard rational approximations all have breaks as follows in the positive quantile region $Z\geq 0, 0.5 \leq u < 1$:
\begin{itemize}
\item Wichura's AS241 \cite{wichura}: two breaks, at $u=0.925$ and $u=1-e^{-25}$.
\item Moro \cite{moro}: breaks at $u=0.92$
\item Acklam Level 1\cite{acklam}:  breaks at $u=0.97575$;
\item Giles' quantile: breaks at $u = 0.9983$
\end{itemize}
Wichura's model is double precision, as is the iterated Acklam (Level 2) model. 

How can we avoid the break, at least for most practical computations? The first thing to point out is that the ``break'' at $u=1/2$ is fictitious in practical applications. It is more sensible to work on a half region, e.g. $0.5\leq u<1$, and output both $Z=\Phi^{-1}(u)$ and $-Z$ for simulation purposes, i.e. always work antithetically. So we focus on the real breaks as in the list above. This break arises in standard approaches due to the fact that the standard quantile $\Phi^{-1}(u)$ has rather a split personality - it is slowly varying in the central region where $u$ is between a half and about $0.9$, and then diverges to infinity as $u \rightarrow 1_-$. This is shown in Fig. 3.
\begin{figure}[hbt]
\centering
\includegraphics[scale=0.8]{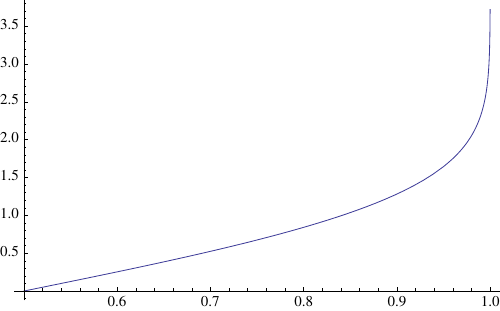}\label{normquant}
\caption{The normal quantile in standard coordinates}
\end{figure}
If we work in an exponential base the situation changes. The function $Q(v) = \Phi^{-1}(1-1/2 e^{-v})$ is shown in Fig. 4 for the region $0\leq v \leq 37$.
\begin{figure}[hbt]
\centering
\includegraphics[scale=0.8]{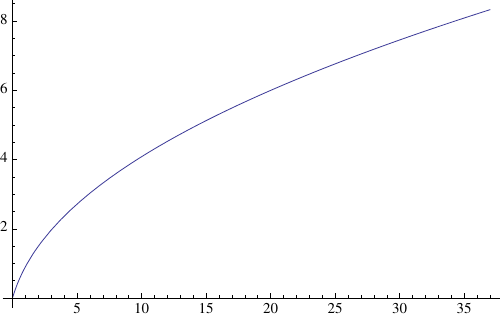}\label{expton}
\caption{The normal quantile in exponential coordinates}
\end{figure}
This function now has a much simpler quality and we can aim to build a single useful rational  approximation. It is then a matter of picking a target range and precision for the desired result. In Fig. 4 we have plotted the function in the range $0\leq v \leq 37$, showing the single character of the function, that will allow a single rational approximation to be made. 

For precision we shall require relative error below $\epsilon_{Rf} =  2^{-25} = 2.98 \times 10^{-8}$, and will consider rational approximations designed to work on a target interval down to at least $10^{-10}$. This was explored using the high-precision arithmetic of Mathematica to work out the normal quantile deep into the tail, and the function {\tt MiniMaxApproximation} to create the rational approximation. The function actually approximated was
\begin{equation}
\frac{Q(v)}{v} = \frac{1}{v} \Phi^{-1}(1-1/2 e^{-v}) 
\end{equation}
and the power series for $Q$ was used in a small neighbourhood of the origin to allow  {\tt MiniMaxApproximation} to work preserving precision near the origin, where $Q(0)=0$. The settings employed for the computation were
\begin{itemize}
\item {\tt Brake -> {20, 20}};
\item {\tt WorkingPrecision -> 20};
\item {\tt MaxIterations -> 400};
\end{itemize}

\subsection{A new normal quantile for float applications}
A rational approximation of degree $(5,6)$ was found with the desired precision. The relative error is shown in Fig. 5 and is less than $2.98 \times 10^{-8}$ on $0\leq v \leq 25$, corresponding to $u$ differing from zero by $2.22 \times 10^{-10}$.
\begin{figure}[hbt]
\centering
\includegraphics[scale=0.8]{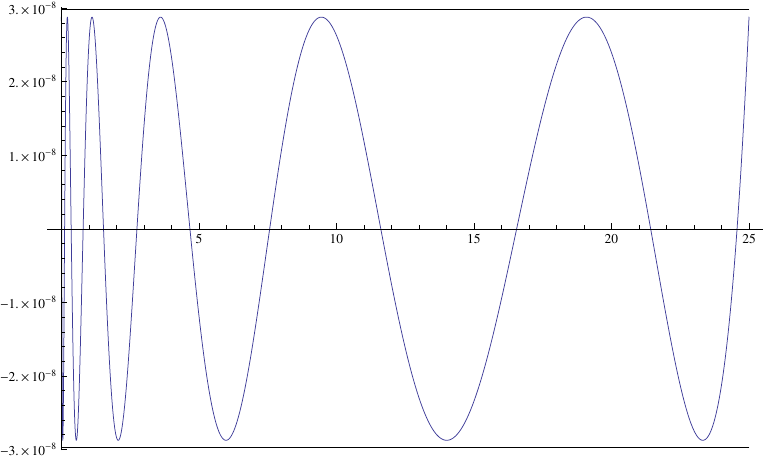}\label{relerr}
\caption{Precision of $(5,6)$ exponential-normal quantile on $[0,25]$.}
\end{figure}
The resulting form for $Q(v)$ is as follows
\begin{equation}
Q(v) = v*p(v)/q(v)
\end{equation}
where $p$ and $q$ are polynomials of degree $5$ and $6$ respectively, with nested C-forms as follows, where we produce the higher-precision output generated by Mathematica. The numerator $p$ is
\begin{verbatim}
1.2533141012558299407 + v*(2.4101601285733391215 
+ v*(1.3348090307272045436 + v*(0.23753954196273241709 
+ (0.011900603295838260268 + 0.00011051591117060895699*v)*v))))                   \end{verbatim}
and the denominator $q$ is
\begin{verbatim}
1 + v*(2.4230267574304831865 + v*(1.8481138350821456213 
+ v*(0.50950202270351517687  + v*(0.046292707412622896113 
+ (0.0010579909915338770381 + 2.5996479253181457637e-6*v)*v)))))
\end{verbatim}
For completeness, an algorithm for normal samples based on this is (in the first two steps we give in brackets the better form using a reflection and scaling to simplify the first part and avoid precision reduction near unity):
\begin{itemize}
\item sample $u$ in $1/2\leq u<1$(or, better, $0<u<1$);
\item evaluate $v =-\log[2(1-u)] $, (then, better, $v =-\log[u] $);
\item evaluate $Z = Q(v)$ with $Q$ given by the rational approximation;
\item output the antithetic pair ${Z, -Z}$.
\end{itemize}
If an exponential base is used we are essentially employing the last two steps. This is what should be done ideally to get maximum precision. A quantile for managing the entire range $(0,1)$ is given in CUDA kernel form in Appendix A. It is that model that is investigated in the next sections with regard to theoretical precision, realized precision and speed.

\subsection{Theoretical precision analysis}
The error characteristics of the two new formula is shown in the Figures 6 and 7.
\begin{figure}[hbt]
\centering
\includegraphics[scale=0.6]{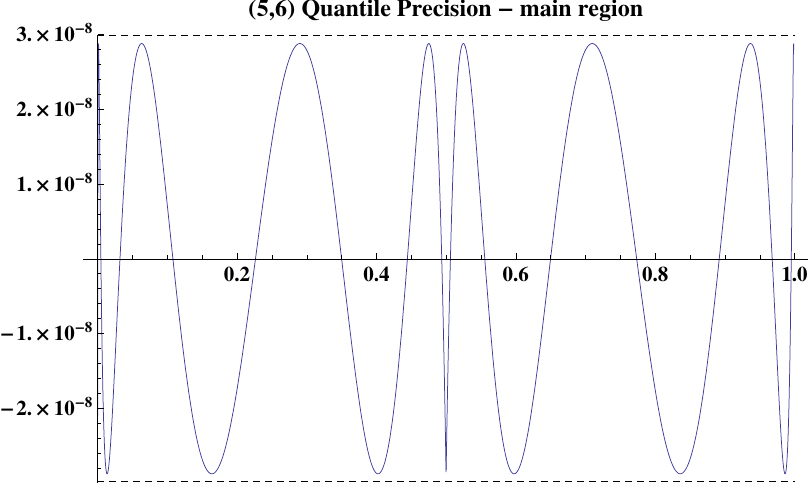}\label{newprec}
\caption{Theoretical relative precision of new branchless quantile - main region}
\end{figure}
\begin{figure}[hbt]
\centering
\includegraphics[scale=0.65]{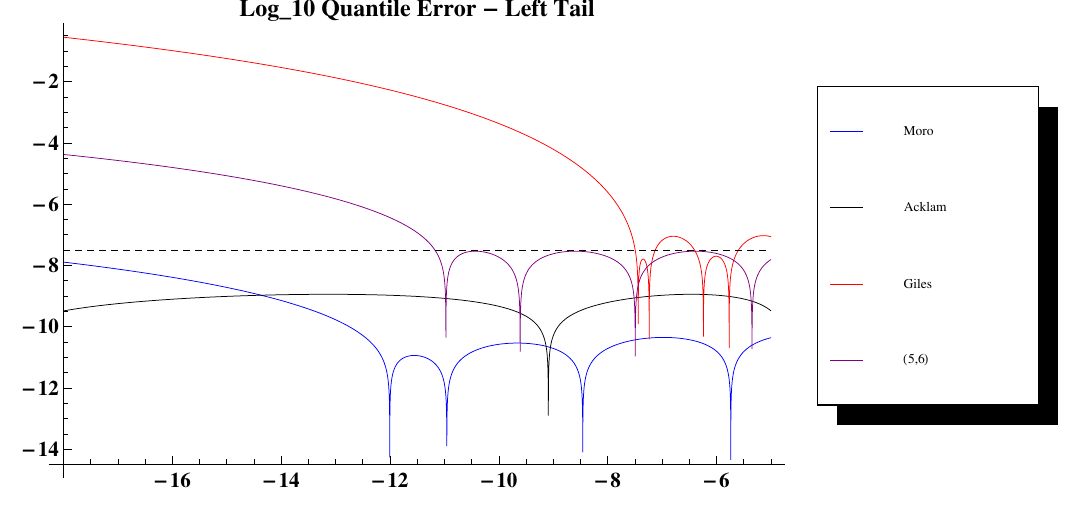}\label{newtail}
\caption{Theoretical relative precision of new quantiles - tail region}
\end{figure}
Fig. 6 illustrates the fact that we have satisfied the precision goal in each case for the main region. The behaviour in the deep tail is shown in Fig. 7. This illustrates the depth of precise tail penetration by the two algorithms. Our view was that $10^{-10}$ is appropriate for Monte Carlo simulations with about $10^8$ samples.  The {\it Mathematica} notebook used to generate such schemes can be obtained on request from WS, and it is not difficult to generate branchless schemes valid over still larger regions. 

In reality these theoretical precisions computed in full precision will not be reproduced when passed through CUDA, or C++ in {\it float} mode. However, we see a significant improvement in the number of significant figures for numbers of order $u= 10^{-7}$. A number of sample computations in CUDA revealed that these schemes were generating {\it realized} precision of order $3\times 10^{-7}$ in contrast to about $1\times 10^{-3}$ from the Giles formula, which has similar characteristics to the CUDA 4 library. 

\subsection{Realized precision analysis}
\begin{figure}[hbt]
\centering
\includegraphics[scale=0.5]{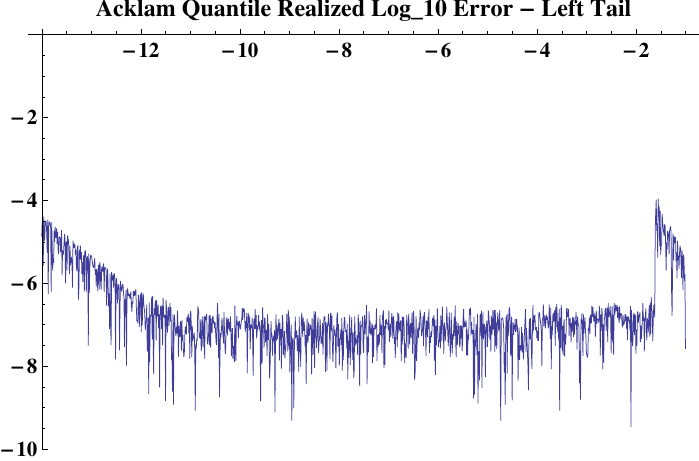}\label{ackrealprec}
\includegraphics[scale=0.5]{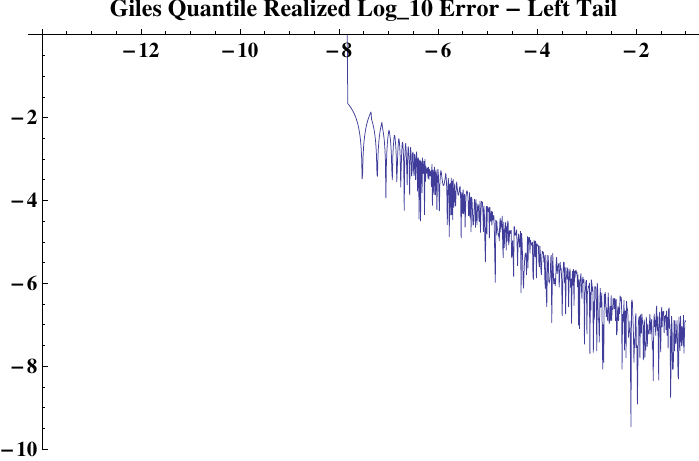}\label{gilesrealprec}
\bigskip
\includegraphics[scale=0.5]{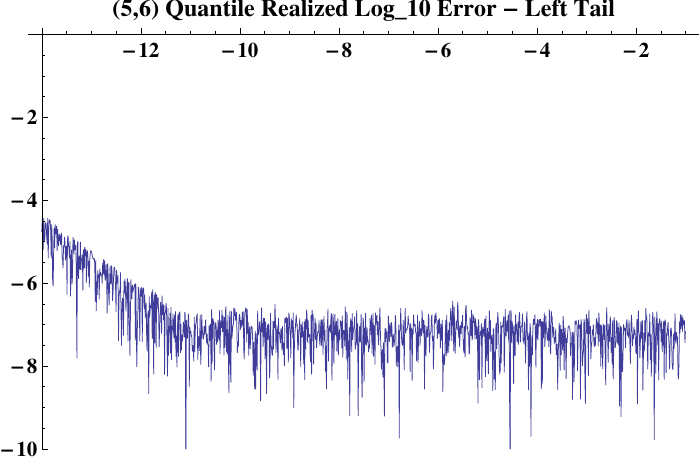}\label{newrealprec}
\caption{Realized relative precision of various quantile in float: left tail}
\end{figure}
To explore this reality we utilized the GPU-link capabilities built in to Mathematica version 8 and passed tail data to the GPU, ran the kernel forms of the Acklam, Giles and $(5,6)$ formulae and then passed the results back to Mathematica. The results are shown in Figure 8. The new formula preserves a precision better than about $3.1 \times 10^{-7}$ down to $u=10^{-11}$. We should also point out that none of these {\it float} formulae can deal with the right hand region effectively. The value of machine epsilon significantly limits the generation of extreme positive values, which is an argument for moving to {\it double} precision.

\subsection{Timing comparisons} 
In our earlier study we compared Acklam's quantile with the codes ICNDfloat1  and ICNDfloat2 and found that Acklam's quantile remains superior on a CPU, but was improved on by the branchless schemes on a GPU. On balance the avoidance of any ``special function'' calls means that on a CPU the central simple scheme wins out, but that on a GPU the application of both square roots and logs in the tail algorithm slows down the code rather dramatically.

Here we will re-analyse the GPU efficiency of the Acklam and Giles models compared to our new $(5,6)$ rational model.  For a proper GPU analysis the code was ported initially to a GTX285 GPU, and followed up with re-computation on a GTX480, Quadro 4000, Tesla C2050 and re-run in the same way. Timings for the quantile call were as given in Table 1.
\begin{table}[hbt]
\begin{center}
\begin{tabular}{|l|l|l|l|l|} \hline
Algorithm &  Timing $t[ms]$ &$t[ms]$ &$t[ms]$ & $t[ms]$ \\ \hline
GPU &  GTX285 & Quadro 4000 & GTX 480 & C2050\\ \hline
Acklam  &$197.0$ & $699.05$& $211.8$ & $279.8$  \\ \hline
Giles  &$104.3$  &$222.56$ & $84.6$ & $114.0$\\ \hline
branchless (5,6)  &$114.2$&$230.07$ & $85.7$& $114.5$\\ \hline
\end{tabular}
\end{center}
\caption{Single precision timings for various normal quantiles on GPU}
\end{table}
The benefit of working in branchless form is now clear. The improvement over the older Acklam form, can make a difference, especially if one is solving an SDE via many calls to a normal random variable prior to evaluating a payoff. Users of GPUs must make a judgement as to the relative merits of the Giles scheme vs our own proposals. In our view a substantial increase in tail accuracy justifies a very small decrease in speed, which is only about 0.5\% on a modern professional class GPU. We also point out that our own computations are based on a supremum norm in the relative error - it is possible that the lack of tail precision in the Giles quantile may be due to the use of an $L_2$ error norm that does not provide control over a tail region that is small in that norm. Our understanding is that the algorithm for {\it erfinv} in the CUDA 4 toolkit is based on the Giles model but uses some additional internal hardware tricks on log etc. to which we do not have access, so some further optimization of the $(5,6)$ model is also likely to be possible - the results here are all based on openly available functions. 

In the full philosophy of this paper, one would of course use an exponential base for many different computations and distributions and possibly pre-store a large set of exponential samples created by efficient methods. The overhead of converting to normal is then the evaluation of a simple rational function and the performance benefits are magnified many times over those given in Table 1.

\section{High precision work}

One can also consider working to double precision on a modern Nvidia GPU. The need for double precision is somewhat contentious in the Monte Carlo framework, but it seems to us that while some computations might be satisfactory in float, one does need to check. While purely additive Monte Carlo summations are almost certainly be OK, one can never be sure about any computation in float until one has seen that repeating in double gives consistent answers. Furthermore, the commonplace approach to working out mathematical derivatives by differencing can generate a substantial precision loss, as it involves an essential subtraction. These are well-established concerns. To these we add the issues associated with generate regions in the right tail with $u$ close to unity, for which the float form of machine epsilon is a killer issue. In our own experiments the right tail no longer has a proper distribution of values but instead has a concentration of particular values of
\begin{equation}
\Phi^{-1}(1 - 2^{-24}) = 5.2497\ ,\  \Phi^{-1}(1 - 2^{-23}) = 5.16658\
\end{equation}
and no higher. Thus the sample is thereby distorted and capped at a low value. While this problem persists in double precision, the limiting levels (as is already noted by Acklam) are at
 \begin{equation}
\Phi^{-1}(1 - 2^{-54}) = 8.29236\ ,\  \Phi^{-1}(1 - 2^{-53}) = 8.20954\
\end{equation}
which are unlikely to be achieved in a simulation of forseeable size.

The next matter to establish is the quality of standard methods. There are two well-established candidates. These are
\begin{enumerate}
\item Wichura's AS241;
\item The {\it refined Acklam} method, where the level one approximation is fed once through a Newton-Raphson-Halley method.
\end{enumerate}
How do we do a quality check on such high precision methods? We will use the {\it Mathematica} function {\tt InverseErf} as our benchmark. However, this will not be done blindly on the assumption that it is necessarily correct. The quantile based on this has been independently assessed against the known exact solution for the Gaussian quantile developed by Steinbrecher and Shaw \cite{qmone} that is known in series form. The formula for this in a computation-suitable representation is also available at {\url http://en.wikipedia.org/wiki/Probit} and as a series has been coded up both in {\it Mathematica} and quadruple-precision FORTRAN based on the Absoft compiler. Based on these three implementations various cross-verifications have been carried out. For example, the quad-precision FORTRAN code that agrees with {\it Mathematica's} internal {\tt InverseErf} to a precision of better than $10^{-29}$ on the interval $[e, 1-e]$, with $e=0.0007$.  Near the centre of the unit interval the truncated series written in {\it Mathematica} agrees with {\tt InverseErf} to much better than quad precision.  So we have considerable confidence that our benchmark is precise enough for any double-precision evaluation. 

In the context of double precision we will restrict attention to realized precision, and this has been evaluated directly using Mathematica 8's capability to embed CUDA kernels. We send a controlled sequence of double-precision numbers in $(0,1)$ to a kernel of interest, send the results back and compare with our high-precision benchmark. First, in Fig. 9 we show the left-tail outcome for doing this for a CUDA implementation of AS241. This confirms that double precision is realized consistently throughout the region of interest. AS241, in common with every scheme, suffers a small loss of realized precision in the neighbourhood of $u=0.5$, where the quantile is zero, and this is shown by the small blip at the right. This is not a matter of concern especially as the absolute precision in a neighbourhood of $u=0.5$ is very good. This plot, in common with the others, shows $-\log_{10}(approx/exact-1)$ as a function of $-\log_{10}(u)$, and the vertical line shows the range bound set by machine epsilon. In practice real Monte Carlo simulations are only likely involve data well to the right of this line.
\begin{figure}[h]
\centering
\includegraphics[scale=0.7]{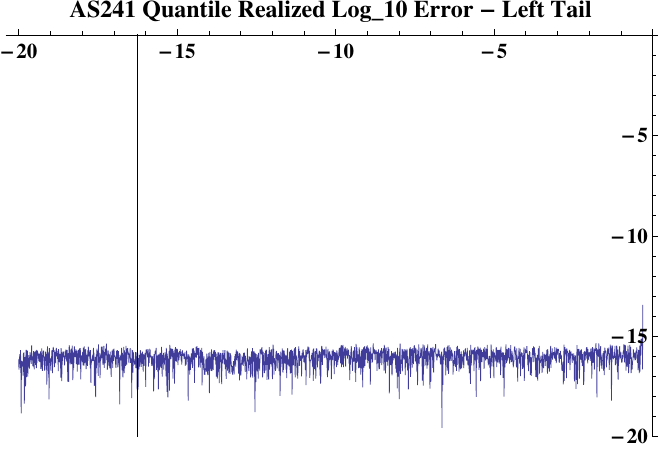}\label{as241}
\caption{Precision of AS241 CUDA kernel in left tail}
\end{figure}
The corresponding error plot for the {\it iterated} Acklam form is shown in Fig. 10. There is, in this implementation, a small glitch in precision for values of $u$ below $10^{-5}$, but the overall levels are very good.\footnote{This might, of course, be a bug in our implementation in CUDA, but we have been unable to find an error, and a corresponding implementation in high precision in Mathematica produces a different manifestation of precision loss.}
\begin{figure}[h]
\centering
\includegraphics[scale=0.7]{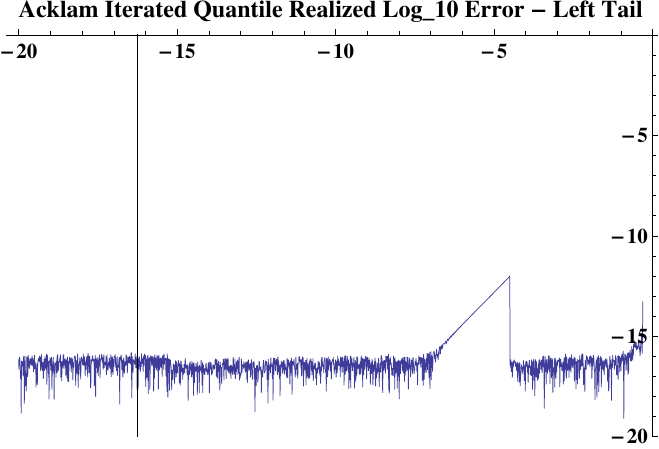}\label{refackmiddle}
\caption{Precision of iterated Acklam in left tail}
\end{figure}

\subsection{Double precision quantiles for GPU work}
In order to produce a faster double-precision quantile for GPU work, we can consider either
\begin{itemize}
\item aiming to be essentially branchless using a single formula over the entire range;
\item having a low warp divergence probability but with a faster formula.
\end{itemize}
In our earlier study \cite{sbarxiv} we produced an essentially branchless formula based on an exponential transformation, and Giles has also generated a double precision analogue in his approach based on a branching formula with a low warp divergence probability. 

First the Giles formula is evaluated on the same basis as AS241 and the iterated Acklam model, and the results are shown in Fig. 11. This, in our view, suffers from the same precision degradation as the float model, and again we suspect that this is due to the use of an $L_2$ norm rather than a supremum norm.

\begin{figure}[h]
\centering
\includegraphics[scale=0.7]{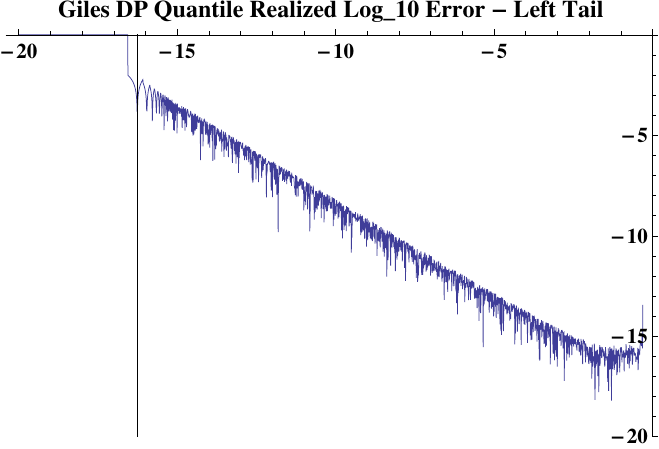}\label{gilesdp}
\caption{Precision of double precision Giles (CUDA 4 similar) model in left tail}
\end{figure}

The production of suitable approximations to compete with these methods follows the same path as considered for the float case, except that now
\begin{itemize}
\item the target for relative precision is now to be less than $\epsilon_{Rd} = 5.55 \times 10^{-17}$
\item the domain is $(\epsilon_{Rd}, 1-\epsilon_{Rd})$.
\end{itemize}

\subsection{A branchless solution}
\begin{figure}[h]
\centering
\includegraphics[scale=0.7]{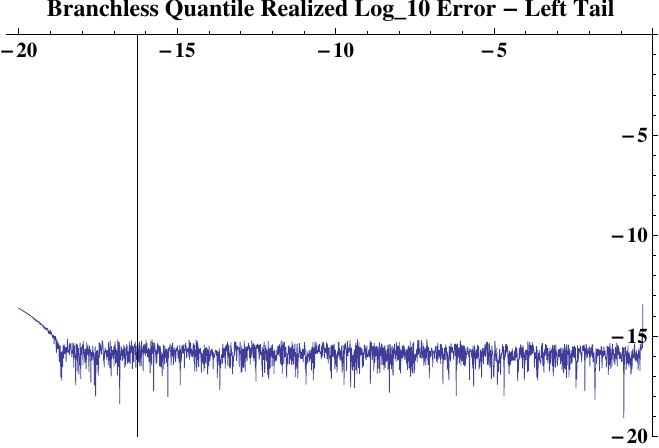}\label{branchlessdp}
\caption{Precision of branchless model in left tail}
\end{figure}
In order to achieve the desired relative precision on the given domain, we first considered a branchless solution based again on the simple exponential distribution. With the change of variables $v = -\log(2*u)$ on the left region $0<u<1/2$, and  $v = -\log(2*(1-u))$ on the right region $1/2<u<1$ it is straightforward to construct a rational approximation of degree $(13,13)$ satisfying an relative theoretical error bound of less than $5.4 \times 10^{-17}$ on an extension of the target domain up to $v = 42$, corresponding to a quantile point of $u = 2.875 \times 10^{-19}$. This is essentially a small variation of our previous double precision model given in \cite{sbarxiv}, also based on a $(13,13)$ model, adapted to target the bound $\epsilon_{Rd}$ within the largest domain possible. The kernel is shown in Appendix B. The realized precision of this model is shown in Fig. 12.

\subsection{Filtering on the t2-distribution}
An interesting feature of some GPUs is that they involved an unusually efficient implementation of the reciprocal square root function
\begin{equation}
{\rm rsqrt(x)} \equiv \frac{1}{\sqrt{x}}
\end{equation}
and in double precision this is {\it much} faster than the logarithm function. This suggests we explore the sequence in which a uniform random variable is mapped to a suitable sample of the Student t distribution with {\it two} degrees of freedom, and then this is re-mapped by appropriate solution of the recycling ODE. Specifically, we will consider an intermediate random variable that is of the form
\begin{equation}
V = \frac{T_2}{\sqrt{2}}
\end{equation}
The quantile function for this random variable is just
\begin{equation}
Q_v(u) = \frac{(u - 1/2)}{\sqrt{u - u*u}}
\end{equation}
and the mapping from $V$ to a normal random variable is the composition of the CDF of $V$ with probit:
\begin{equation}
\Psi(v) = \Phi^{-1}\biggl( \frac{1}{2} \biggl[1 + \frac{v}{\sqrt{1 + v^2}}\biggr]\biggr)
\end{equation}
We have experimented with the rational approximation of $\Psi$ and while it is difficult to give an economical branchless solution, it serves as the basis for a very fast branching solution but with low warp divergence. Specifically, we have found a rational approximation of degree $(14,15)$ that satisfies the error bound for $0 \leq v \leq 15.5$. Furthermore,
\begin{equation}
\Psi(15.5) = 3.07933
\end{equation}
for which $u = 0.998963$, giving a low warp divergence probability, if implemented as a standard "IF" statement. This therefore provides the basis for a hybrid model in which we put back a branch point, but deep into the tail. In the central region the computation is done by using the $T_2$ as an intermediate distribution, and in the tail the computation is done using the already computed branchless solution. The precision characteristics are indistinguishable from the branchless case and are shown in Fig. 13. 

There is one final point about the way we have implemented the branching for this case, as we do not have warp divergence at all. The algorithm implements the central $T_2$ filter if all the threads the warp should follow the central path, otherwise all the threads fall back to our branchless model. This is accomplished in the code by warp voting using the $all()$ command.
\begin{figure}[h]
\centering
\includegraphics[scale=0.7]{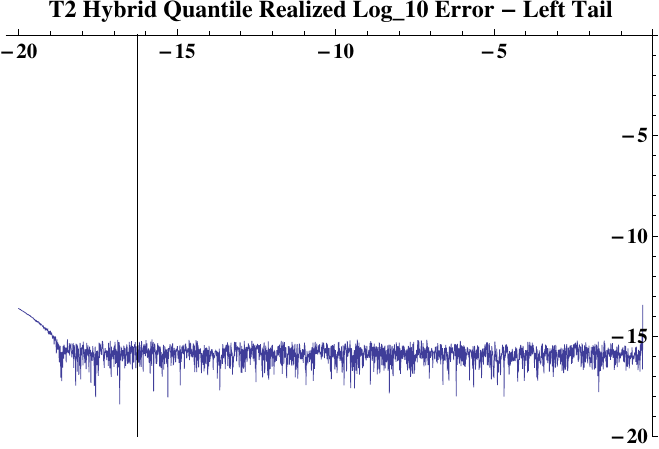}\label{hybrid}
\caption{Precision of hybrid model in left tail}
\end{figure}

\subsection{Timing comparisons}

The comparisons on a modern GPU are very interesting. We have completed a comparison of seven algorithms on four GPUS. The four algorithms are:
\begin{itemize}
\item Acklam's one-pass method, but coded in double precision (DP);
\item The Acklam-Lea DP method;
\item AS241 (Burkardt code as of early 2009);
\item the CUDA 4 erfinv model;
\item Giles double precision erfinv model;
\item Our latest pure breakless model;
\item Our $T_2$ exponential hybrid. 
\end{itemize}
Our understanding is that the CUDA and Giles models are different for double precision, though we have found them to have similar precision characteristics. The four GPUs considered are
\begin{itemize}
\item GTX 285;
\item Quadro 4000;
\item GTX 480;
\item Tesla C2050.
\end{itemize}
The timings in $ms$ for the test program are reported in Table 2 with compilation for compute capability 1.3 in the case of the 285, and 2.0 in the case of the Quadro 4000, GTX 480 and Tesla C2050.

\begin{table}[hbt]
\begin{center}
\begin{tabular}{|l|r|r|r|r|} \hline
Algorithm &Timing $t[ms]$&  Timing $t[ms]$&  Timing $t[ms]$&  Timing $t[ms]$ \\ \hline
\hspace{1.5in}GPU:   & GTX 285 & Quadro 4000 & GTX 480 & Tesla C2050 \\ \hline
Acklam non-iterated as double  &$2391.97$&$1278.25$&$949.46$&$611.44$ \\ \hline
Acklam iterated &$5821.62$&$3133.62$  &$2305.16$&$1488.53$\\ \hline
AS241  &$3014.99$&$1675.05$&$1285.20$&$797.88$  \\ \hline
CUDA 4 DP  &$3541.24$&$1950.91$&$1441.11$&$930.36$  \\ \hline
Giles DP &$1594.69$&$1016.50$  &$811.51$&$486.49$\\ \hline
Breakless (Appendix B) &$2013.04$&$1178.17$  &$889.35$&$562.82$\\ \hline
Hybrid $T_2$ (Appendix C) &${\bf 1551.58}$&${\bf 933.01}$  &${\bf 607.20}$&${\bf 448.13}$\\ \hline
\end{tabular}
\end{center}
\caption{Double precision timings for various normal quantiles on GPU}
\end{table}
The use of the $T_2$ as an intermediate distribution results in a clear win for this method on performance grounds, and given that it also provides clear double precision accuracy over the range of interest for Monte Carlo work, we argue that it is a good candidate for the preferred method on GPUs.  We note that this choice seems best reserved for double precision work - the advantage arises from the relatively fast speed of the inverse square root operation compared to the logarithm in double precision under CUDA. We do not see this advantage in float arithmetic. We also note that on GTX cards, it is possible to go slightly faster still\footnote{Requests for code implementing this idea can be made to the authors} using a hybrid with more branches and optimized $T_2$ intermediates for a near tail, but these offer no benefit on the C2050 Tesla. Finally our latest breakless method comes in 3rd behind the hybrid and Giles approaches, but has the merit of robustness with respect to any increase in the number of threads per warp.  We also briefly explored polynomial representations of the mapping in the central region in order to avoid a costly divide, but found it hard to get close to the target supremum error bound. An example computation with degree 32 over the target range $[0,15.5]$ got us down to only $O(10^{-7})$.

\section{Hyperbolic and Variance Gamma - some groundwork}
In this section we move to other distributions of interest to finance. First we consider the {\it hyperbolic} distribution, and then the variance gamma. These will have in common a non-normal base distribution, and will illustrate the use of a 2-sided exponential base instead.

\subsection{Hyperbolic quantile from exponential base}

This was originally motivated by Bagnold's classic study of sand \cite{bagnold}, and was given a clear mathematical description by Barndorff-Nielsen \cite{bnhyp}, who also generalized it. The applications to finance have been explored Eberlein and Keller \cite{eberlein}. A direct treatment of the quantile function for the symmetric case has been given by Xiong \cite{ying}. He we shall explore the conversion of samples from a suitable exponential distribution to samples from the hyperbolic. Hyperbolic distributions can of course be sampled as random mixtures of a normal distribution. Our method facilitates the use of hyperbolic marginals coupled to an arbitrary copula, and and this example also illustrates how cleanly the choice of a suitable base simplifies the computations of the quantile - the exponential base regularizes the tail in an elegant way.

The probability density function is known explicitly as
\begin{equation}
f(x,\alpha,\beta,\delta,\mu) = \frac{\gamma}{2\alpha \delta K_1(\delta \gamma)}\exp\{-\alpha\sqrt{\delta^2 + (x-\mu)^2} + \beta(x-\mu)\}
\end{equation}
where $\gamma = \sqrt{\alpha^2-\beta^2}$, with $|\beta| <\alpha $. In what follows we shall translate the origin so that $\mu=0$, with density 
\begin{equation}
f(x,\alpha,\beta,\delta) = \frac{\gamma}{2\alpha \delta K_1(\delta \gamma)}\exp\{-\alpha\sqrt{\delta^2 + x^2} + \beta x\}
\end{equation}
The $H$-function for the target distribution is given by the negative of the logarithmic derivative:
\begin{equation}
H(x) = -\frac{d\ }{dx} \log f(x,\alpha,\beta,\delta)  = \frac{\alpha x}{\sqrt{\delta^2+x^2}}-\beta
\end{equation}
and it is evident that for large $x$,
\begin{equation}
H(x) \sim {\rm sign}(x) \alpha - \beta = \pm \alpha - \beta\ .
\end{equation}
Bearing in mind that the exponential distribution is characterized by a constant $H$-function, we will use a pair of exponential distributions for the base case. When we solve the resulting recycling equations, it will simplify matters to set $Q(0)=0$ for the boundary conditions, and in order to get the proportion of the random variables that are positive and negative correct with this choice, we let
\begin{equation}
\begin{split}
p_+ & = \int_0^{\infty}\!dx \frac{\gamma}{2\alpha \delta K_1(\delta \gamma)}\exp\{-\alpha\sqrt{\delta^2 + x^2} + \beta x\}\\
p_- & = \int_{-\infty}^0\!dx \frac{\gamma}{2\alpha \delta K_1(\delta \gamma)}\exp\{-\alpha\sqrt{\delta^2 + x^2} + \beta x\}
\end{split}
\end{equation}
so clearly $p_+ + p_-=1$. We choose the base distribution to have the form
\begin{equation}
f_0(x) = \begin{cases} 
p_+(\alpha-\beta) e^{-(\alpha-\beta)x}& \text{if $x>0$,}
\\
p_- (\alpha+\beta) e^{(\alpha+\beta)x}& \text{if $x<0$,}
\end{cases}
\end{equation}
The quantile function for sampling from $f_0$ has the trivial form:
\begin{equation}
v = Q_0(u) = \begin{cases} 
\frac{1}{\alpha+\beta}\log(u/p_-)& \text{if $u<p_-$,}
\\
\frac{-1}{\alpha-\beta}\log((1-u)/p_+)& \text{if $u>p_-$,}
\end{cases}
\end{equation}
So samples from the base can be made easily. To convert them into samples from the hyperbolic we solve a {\it left} and {\it right} differential equation. The right problem is of the form
\begin{equation}
\frac{d^2 Q}{dv^2} + (\alpha-\beta)\frac{dQ}{dv} = \biggl(\frac{\alpha Q}{\sqrt{\delta^2+Q^2}}-\beta  \biggr) \biggl(\frac{dQ}{dv} \biggr)^2
\end{equation}
on $v>0$ with the initial condition $Q(0)=0$ and
\begin{equation}
\frac{d Q}{dv}|_{v=0} =\frac{Q'(p_-)}{Q_0'(p_-)} = \frac{f_0(0_+)}{f(0)} = p_+(\alpha-\beta)2 \frac{\alpha \delta}{\gamma} K_1(\delta \gamma)e^{\alpha \delta}
\end{equation}
The left problem is 
\begin{equation}
\frac{d^2 Q}{dv^2} - (\alpha+\beta)\frac{dQ}{dv} = \biggl(\frac{\alpha Q}{\sqrt{\delta^2+Q^2}}-\beta  \biggr) \biggl(\frac{dQ}{dv} \biggr)^2
\end{equation}
on $v<0$ with the initial condition $Q(0)=0$ and
\begin{equation}
\frac{d Q}{dv}|_{v=0} =\frac{Q'(p_-)}{Q_0'(p_-)} = \frac{f_0(0_-)}{f(0)} = p_-(\alpha+\beta)2 \frac{\alpha \delta}{\gamma} K_1(\delta \gamma)e^{\alpha \delta}
\end{equation}
The solution to this differential system is readily visualized. If we use a sixth-order explicit RK method as before, with parameters $\alpha=1=\delta, \beta=0$ for illustration, the result for the recycling mapping is show below, together with the identity map (diagonal line). 
\begin{figure}[hbt]
\centering
\includegraphics[scale=0.8]{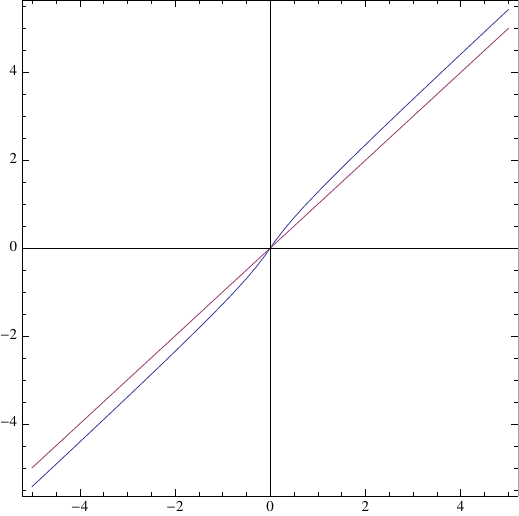}
\caption{Plot of the mapping for conversion of exponential to hyperbolic}
\end{figure}

\subsection{VG quantile from exponential base}
The variance-gamma density was introduced by Madan and Seneta \cite{madan} as a model for share market returns. The density is given, for $\lambda>0, \alpha>0, |\beta|<\alpha$, by
\begin{equation}
\frac{e^{\beta x} |x|^{\lambda -\frac{1}{2}}
   \left(\alpha ^2-\beta ^2\right)^{\lambda } K_{\lambda
   -\frac{1}{2}}(\alpha  |x|)}{(2\alpha)^{\lambda-1/2}\sqrt{\pi } \Gamma (\lambda )}
\end{equation}
In the region $x>0$ the $H$-function is given by  
\begin{equation}
H(x) = -\frac{d\ }{dx} \log (f)  = \frac{\alpha  K_{\lambda-3/2}(\alpha x
   )}{K_{\lambda-1/2}(\alpha x)}-\beta \sim (\alpha -\beta )+\frac{1-\lambda
   }{x}+O\left(\left(\frac{1}{x}\right)^2\right)
   \end{equation}
   In the region $x<0$ the $H$-function is given by  
\begin{equation}
H(x) = -\frac{d\ }{dx} \log (f)  =- \frac{\alpha  K_{\lambda-3/2}(-\alpha x
   )}{K_{\lambda-1/2}(-\alpha x)}-\beta \sim -(\alpha +\beta )+\frac{1-\lambda
   }{x}+O\left(\left(\frac{1}{x}\right)^2\right)
   \end{equation}
These asymptotic relationships suggest that the VG model may be treated in a similar way to the hyperbolic case, as the asymptotics are closely related with a good match to the exponential base.  This time the probabilities $p_{\pm}$ are given by

\begin{equation}
\begin{split}
p_+ & =  \frac{\left(\alpha ^2-\beta ^2\right)^{\lambda } }{(2\alpha)^{\lambda-1/2}\sqrt{\pi } \Gamma (\lambda )} \int_0^{\infty}\!dx e^{\beta x} x^{\lambda -\frac{1}{2}}
K_{\lambda
   -\frac{1}{2}}(\alpha x )\\
   &=\frac{2^{2 \lambda -1} \left(\frac{\alpha +\beta }{\alpha
   -\beta }\right)^{\lambda } \Gamma \left(\lambda
   +\frac{1}{2}\right) \, _2F_1\left(2 \lambda ,\lambda
   ;\lambda +1;\frac{\alpha +\beta }{\beta -\alpha
   }\right)}{\sqrt{\pi } \Gamma (\lambda +1)}, \\
p_- & =  \frac{\left(\alpha ^2-\beta ^2\right)^{\lambda } }{(2\alpha)^{\lambda-1/2}\sqrt{\pi } \Gamma (\lambda )} \int_0^{\infty}\!dx e^{-\beta x} x^{\lambda -\frac{1}{2}}
K_{\lambda
   -\frac{1}{2}}(\alpha x )\\
   &=\frac{2^{2 \lambda -1} \left(\frac{\alpha -\beta }{\alpha
   +\beta }\right)^{\lambda } \Gamma \left(\lambda
   +\frac{1}{2}\right) \, _2F_1\left(2 \lambda ,\lambda
   ;\lambda +1;\frac{\beta -\alpha }{\alpha +\beta
   }\right)}{\sqrt{\pi } \Gamma (\lambda +1)},\end{split}
\end{equation}
where we have used identity 6.621.3 from \cite{amsteg} to evaluate the integrals giving the probabilities that the VG random variables is positive or negative. It is easily checked that if $\beta=0$ then $p_+=p_-=1/2$.

The difference between VG and hyperbolic is that in the case of VG the details of what has to be done are sensitive to the value of $\lambda$. First, we note that if $\lambda=1$ the VG model is trivial as it is identical to the base, so that $Q(v)\equiv v$. If $\lambda >1$ matters remain reasonably straightforward, as both $f$ and $H$ exist at $v=0$, with $H(0)=0$. The recycling ODE may be solved as before, though many steps may be needed near $v=0$ if $\lambda$ remains close to and just above $1$. When $0<\lambda<1$ matters are more complicated, as then $H(0)$ is divergent, and furthermore the density becomes singular in the range $0< \lambda \leq 1/2$. The density has a log divergence when $\lambda=1/2$, and otherwise diverges as $x^{2\lambda-1}$. All of these issues may in principle be addressed by doing analytical estimates in a small neighbourhood of the origin and starting the numerical treatment at a small distance from the origin - as noted several different cases must be considered and full details will be given elsewhere.

\section{Conclusions}
This paper has developed two related  threads of thought. First, risk simulations depend critically on having a realistic (fat-tailed) model of asset returns. The methods developed here allow traditional Gaussian samples to be converted to other distributions via the application to the samples of the solution of a recycling differential equation. The recycling differential equation is the ODE for transforming samples from a density $f_0$ to a density $f_1$, and is
$$
\frac{d^2 Q(v)}{dv^2}+H_0(v)\frac{dQ(v)}{dv} = H_1(Q(v)) \left(\frac{dQ(v)}{dv} \right)^2\ , \label{peafour}
$$
where 
$$H_i = -\frac{d\ }{dx} \log [f_i(x)]$$
We have given an explicit example for the Student t case, where a power series emerges coupled to a tail model. Other more complicated distributions with an explicit density may be handled similarly or numerically, and other base distributions may be treated. In particular we can use changes of variable to construct ``essentially IF-less'' algorithms for objects like the normal quantile. The efficiency of such algorithms in GPU computation is of interest, and the methods introduced here can be considered for other target distributions. In contrast to the normal case, where there are no parameters beyond the translation and scale, we must first solve the RODE with the relevant parameters and then develop a suitable fast approximation.

These methods also simplify the use of a Gaussian or T-copula, since the two steps of mapping to the unit hypercube and back to the marginals may be folded together into one operation, where the solution to the RODE is applied directly in one step. Of course, the methods developed here rely on the ability to compute the logarithmic derivatives of the target and base densities. Where the target density is not known explicitly, but its characteristic function is known, other methods must be used. Investigations of the resulting integro-differential equations will be reported elsewhere.

The second thread of thought concerns the detailed practicalities of building good quantile implementations for Monte Carlo use. We have set out the numerical and computer science issues that are relevant to this in general. Then,  as a particular application of these methods for recycling, we have carried out an extensive re-evaluation of the methods used for computing the normal quantile, especially in a GPU environment.  We have reported  new formulae for the normal quantile in both {\it float} and {\it double} mode. In float mode our approach is competitive on speed with the fast formula developed by Giles but offers, we believe, a substantial improvement in precision. In double precision we have provided a new formula which outperforms all other methods known to us on GPU speed while preserving full double precision on a range more than enough for Monte Carlo simulations.

\section*{Acknowledgments}
TL thanks the Knowledge Transfer Network, EPSRC and the Numerical Algorithms Group Ltd for support under a KTN-EPSRC CASE award. WS wishes to thank I. Buckley, W. Gilchrist, P. J\"{a}ckel, D. Scott, G. Steinbrecher , A. Munir and Y. Xiong for discussions on various aspects of quantile theory. Presentations by C. Albanese, M.Giles and G. Ziegler and the NAG team at a recent London workshop on GPU computing \cite{workshop} stimulated the development of the essentially ``IF-less'' normal quantile. A. Munir provided assistance in producing further Windows binaries for the 1.3 CUDA architecture. Clarification of precision issues from several members of the Numerical Algorithms Group were invaluable, and we are also grateful to NAG for access to their Tesla system. We are also grateful to Wolfram Research for their creation of the CUDA embedding tools in Mathematica 8 that allowed the precision studies here to be made so easily, and to Tom Wickham-Jones for clarifying some points on double-precision analysis in that environment.

\section*{Appendix A: The (5,6) rational CUDA kernel for float code}
Here is the code for the  (5,6) rational model  in a form suitable for CUDA. The extra significant figures are ignored by the compiler, but we include the full precision form generated in Mathematica for information. The code is written as a sequence of maps $p \rightarrow p*a + b$ in order to make good use of FMA operations. 

\begin{verbatim}
__inline__ __device__ float ws_norminvf(float u)
{ float half_minus_u = 0.5f - u;
  float v, p, q;
  float x = copysignf(2.0f*u, half_minus_u);
  if (half_minus_u < 0.0f) x += 2.0f;
  v = -__logf(x);

  p =       1.1051591117060895699e-4f;
  p = p*v + 0.011900603295838260268f;
  p = p*v + 0.23753954196273241709f;
  p = p*v + 1.3348090307272045436f;
  p = p*v + 2.4101601285733391215f;
  p = p*v + 1.2533141012558299407f;

  q =       2.5996479253181457637e-6f;
  q = q*v + 0.0010579909915338770381f;
  q = q*v + 0.046292707412622896113f;
  q = q*v + 0.50950202270351517687f;
  q = q*v + 1.8481138350821456213f;
  q = q*v + 2.4230267574304831865f;
  q = q*v + 1.0f;
  return -__fdividef(p, q) * copysignf(v, half_minus_u);}
\end{verbatim}

\section*{Appendix B: A double precision branchless quantile kernel}
Here is an optimized double  precision branchless algorithm in a form suitable for CUDA kernel use. The full data supplied by Mathematica's high-precision rational function generator is included.
\begin{verbatim}
extern "C" __global__ void ws_norminv_exp_42(double *in, double *out)
{
  int i = threadIdx.x + blockIdx.x * blockDim.x;
  double u = in[i];
  double half_minus_u = 0.5 - u;
  double v, p, q;
  double x = copysign(2.0*u, half_minus_u);
  if (half_minus_u < 0.0) x += 2.0;
  v = -log(x);

  p =       1.64783242453158904095515084024e-14;
  p = p*v + 5.06687427282961778456165208105e-11;
  p = p*v + 2.10154247206828001641073444523e-8;
  p = p*v + 2.79486316248312621569098418063e-6;
  p = p*v + 0.000158143467460605125860139269297;
  p = p*v + 0.00438343320745866724879101963414;
  p = p*v + 0.0646753575778845943457494008377;
  p = p*v + 0.535690416737220756622791398354;
  p = p*v + 2.57714610175675729492631703269;
  p = p*v + 7.33285309828701618935546741859;
  p = p*v + 12.3353630302640508603664862349;
  p = p*v + 11.9187726041215161859997693572;
  p = p*v + 6.06634828333794870534194478115;
  p = p*v + 1.25331413731550018371372639809;

  q =       9.3774528584890379942301072137e-13;
  q = q*v + 8.67759442958410980713288964586e-10;
  q = q*v + 1.9465409869330334204439096215e-7;
  q = q*v + 0.0000166601689658474353532677312063;
  q = q*v + 0.00066147322306910897444136114895;
  q = q*v + 0.013581089497310892038923062896;
  q = q*v + 0.154424951968123464901887026825;
  q = q*v + 1.01815001279043960887846096372;
  q = q*v + 4.01114257592029176980269694161;
  q = q*v + 9.58786255809221297975776809938;
  q = q*v + 13.8641781886242409731295280702;
  q = q*v + 11.7514614079486467058484941458;
  q = q*v + 5.34024563572829223828055331064;
  q = q*v + 1.0;

// swap the following comments if using non-Fermi card

  out[i] = p * __drcp_rn(q) * copysign(v, -half_minus_u);
  //out[i] = p / q * copysign(v, -half_minus_u);
}
\end{verbatim}
\newpage
\section*{Appendix C: A double precision optimized quantile kernel}
This is the code for the optimized hybrid using both $T_2$ and exponential intermediates.

\begin{verbatim}
__inline__ __device__ double ws_norminv(double u)
{
  double u_minus_half = u - 0.5;
  double v, p, q;
  v = u_minus_half*rsqrt(u - u*u);
  v = copysign(v, 1.0);

  if (__all(v < 15.5)) // just use primary transformation
  { p =       1.782104085988425639109749e-8;
    p = p*v + 6.440474519924356219069418e-6;
    p = p*v + 0.000482136732375834750227199;
    p = p*v + 0.01232844599180035041777457;
    p = p*v + 0.1339079848194463772055615;
    p = p*v + 0.7321232169482363313048945;
    p = p*v + 2.430915571221008791928114;
    p = p*v + 5.655603740868838565046439;
    p = p*v + 9.907612807645135082109572;
    p = p*v + 13.51564899715023382722088;
    p = p*v + 14.63872140375810008418256;
    p = p*v + 12.41571909215588206897004;
    p = p*v + 8.255845342301247665027723;
    p = p*v + 3.804419247607286580763273;
    p = p*v + 1.253314137315500185908045;

    q =       3.410078388443805543169697e-9;
    q = q*v + 1.488872498545715387659909e-6;
    q = q*v + 0.0001291872317875683976854636;
    q = q*v + 0.00381876799889919727517817;
    q = q*v + 0.04841372227036886168190771;
    q = q*v + 0.3111370832026527448772247;
    q = q*v + 1.203884715056252135700492;
    q = q*v + 3.231436646211214118049417;
    q = q*v + 6.505661571707998298885286;
    q = q*v + 10.26672134308754537045644;
    q = q*v + 12.93267136518991650838875;
    q = q*v + 13.17959450512192709608378;
    q = q*v + 10.62936552707102434538252;
    q = q*v + 6.825412147203414419893086;
    q = q*v + 3.035487380487070955193619;
    q = q*v + 1.0;
  }
  else // fallback to exponential transformation
  {
    double x = copysign(2.0*u, -u_minus_half);
    if (u_minus_half > 0.0) x += 2.0;

    v = -log(x);

    p =       1.64783242453158904095515084024e-14;
    p = p*v + 5.06687427282961778456165208105e-11;
    p = p*v + 2.10154247206828001641073444523e-8;
    p = p*v + 2.79486316248312621569098418063e-6;
    p = p*v + 0.000158143467460605125860139269297;
    p = p*v + 0.00438343320745866724879101963414;
    p = p*v + 0.0646753575778845943457494008377;
    p = p*v + 0.535690416737220756622791398354;
    p = p*v + 2.57714610175675729492631703269;
    p = p*v + 7.33285309828701618935546741859;
    p = p*v + 12.3353630302640508603664862349;
    p = p*v + 11.9187726041215161859997693572;
    p = p*v + 6.06634828333794870534194478115;
    p = p*v + 1.25331413731550018371372639809;

    q =       9.3774528584890379942301072137e-13;
    q = q*v + 8.67759442958410980713288964586e-10;
    q = q*v + 1.9465409869330334204439096215e-7;
    q = q*v + 0.0000166601689658474353532677312063;
    q = q*v + 0.00066147322306910897444136114895;
    q = q*v + 0.013581089497310892038923062896;
    q = q*v + 0.154424951968123464901887026825;
    q = q*v + 1.01815001279043960887846096372;
    q = q*v + 4.01114257592029176980269694161;
    q = q*v + 9.58786255809221297975776809938;
    q = q*v + 13.8641781886242409731295280702;
    q = q*v + 11.7514614079486467058484941458;
    q = q*v + 5.34024563572829223828055331064;
    q = q*v + 1.0;
  }
// swap comments on following lines if on non-Fermi card
  return p * __drcp_rn(q) * copysign(v, u_minus_half);
  //return p / q * copysign(v, u_minus_half);}
\end{verbatim}
These codes and justification of the precision plots in this paper are available at

\url{http://www.homepages.ucl.ac.uk/~ucahwts/quantiles/}
\end{document}